 \documentclass[11pt]{article}

 \pdfoutput=1

 \usepackage[polutonikogreek,english]{babel}

 \usepackage{textgreek} 

\usepackage{bigints}

\usepackage[round]{natbib} 

\usepackage{graphicx,siunitx,xspace}
\usepackage{amsmath}
\usepackage{amssymb}
\usepackage{hyperref}
\usepackage{enumitem}
\usepackage{color}
\usepackage{enumitem}
\usepackage{float}
\usepackage{amsfonts,bm}
\usepackage{epstopdf}
\usepackage[mathscr]{euscript}

  \usepackage[normalem]{ulem}
  \usepackage{textcomp}

\usepackage{dashbox}

\usepackage{cancel}

\usepackage[utf8]{inputenc}
\usepackage[T1]{fontenc}
\usepackage{pifont} 

\usepackage{chemformula}

\usepackage{float}
\usepackage{ifthen}
\begin{document}


\begin{center}
{\bf \Large
An English translation of 3 papers of Max 
\citet[][]{Planck_Verh_dtsch_Phys_Ges_1915a,
Planck_Verh_dtsch_Phys_Ges_1915b,
Planck_Verh_dtsch_Phys_Ges_1915c}:}
\\ \vspace*{2mm}
\hspace*{-0mm}
{\bf \Large (a,\,c)~``\underline{Die Quantenhypothese für Molekeln mit mehreren Freiheitsgraden}\,''}
\\ \vspace*{2mm}
{\bf \Large (b)~``\,\underline{Bemerkung über die Entropiekonstante zweiatomiger Gase}\,'',}
\\ \vspace*{2mm}
{\bf \Large namely:}
\\ \vspace*{2mm}
{\bf \Large ``\,\underline{The quantum hypothesis for molecules with multiple degrees of freedom}\,''}
\\ \vspace*{2mm}
{\bf \Large and: ``\,\underline{Remarks on the entropy constant of diatomic gases}\,'',}
\\ \vspace*{2mm}
{\bf \Large
to provide a readable version of the German contents.
}
\\ \vspace*{2mm}
{\bf \large\color{blue}
Translated by Dr. Hab. Pascal Marquet 
}
\\ \vspace*{2mm}
{\bf\bf\color{blue}  \large Possible contact at: 
    pascalmarquet@yahoo.com}
    \vspace*{1mm}
    \\
{\bf\bf\color{blue} 
    Web Google-sites:
    \url{https://sites.google.com/view/pascal-marquet}
    \\ ArXiv: 
    \url{https://arxiv.org/find/all/1/all:+AND+pascal+marquet/0/1/0/all/0/1}
    \\ Research-Gate:
    \url{https://www.researchgate.net/profile/Pascal-Marquet/research}
}
\\ \vspace*{1mm}
\end{center}

\hspace*{65mm} Version-1 / \today

\vspace*{-2mm} 
\begin{center}
--------------------------------------------------- 
\end{center}
\vspace*{-11mm}

\bibliographystyle{ametsoc2014}
\bibliography{Book_FAQ_Thetas_arXiv}

\vspace*{-2mm} 
\begin{center}
--------------------------------------------------- 
\end{center}
\vspace*{-2mm}

Uncertainties/alternatives in the translation are indicated {\color{blue} (in blue)} with {\it\color{blue} italic terms}, together with some additional footnotes and extra terms  (indicated with {\it\color{blue} P. Marquet} or sometimes in {\it\color{magenta}magenta}). 

I have written {\bf in bold} those parts of the text that deal in particular with the problem of determining {\bf the ``\,zero-point energy\,''} {\it\color{blue}(\,``\,Nullpunktsenergie\,''\,)}.
It is indeed for this aspect that I felt the need to translate the 2 thesis memoirs (1879 for the Doctor dissertation; 1880 for the Habilitation) and the 3 articles (1887 about the conservation of energy; 1916 about the absolute definition of the entropy; 1943 about the discovery of quanta in physics), all written in German by Max Planck.

I have, of course, kept \dashuline{the original text of Planck (in black)} unchanged, while sometimes including {\it\color{blue}\dashuline{additional notes (in blue)}}. 

Do not hesitate to contact me in case of mistakes or any trouble in the English translation from the German text.

\vspace*{-4mm}
\begin{center}
========================================================
\end{center}
\vspace*{-11mm}

  \tableofcontents

\newpage
\begin{center}
\underline{\Large\bf The quantum hypothesis for molecules with multiple degrees of freedom}
\vspace*{0.2mm} 

{\large\bf by Max 
 \citet[5 November][]{Planck_Verh_dtsch_Phys_Ges_1915a}}
\end{center}
\vspace*{-4mm}

\setcounter{section}{-1} 
\setcounter{equation}{0}
\setcounter{figure}{0}  

\section{\underline{-- Planck (1915a) -- Introduction} (p.407-408)}
\label{Section-17a-Introduction}
\vspace*{-2mm}

In the following, as usual, I refer to the number of degrees of freedom of a body as the number of its free coordinates (not counting the associated velocities).
Then an oscillator that oscillates in a straight line has a single degree of freedom, as does a rigid body with a fixed axis of rotation.
In contrast, a rigid body rotating about a fixed point has three degrees of freedom.

The law of uniform energy distribution states that in a system of very many molecules of the same nature (whose energy is composed of the squares of the coordinates and the velocities additively), each degree of freedom has on average the amount of energy $k\:T/2$ or $k\:T$, depending on whether this degree of freedom corresponds only to kinetic or also to potential energy. 
However, as is well known, the condition of uniform energy distribution is by no means always fulfilled in nature.
Rather, as the temperature decreases, the average energy of a degree of freedom decreases much more than it should according to this 
theorem.{\color{blue}\footnote{$\:${\it\color{blue}Here, Planck referred to the fact that the specific heat $c_v(T)$ is more and more different at low temperatures from the monoatomic (ideal gas) limit value $c_v \approx (3/2)\:R$, which is only valid for the higher temperatures / P. Marquet}.}}

The actual connection between energy and temperature, and thus also the amount of specific heat, can be found by using the thermodynamic equation:  
\begin{align}
\frac{d\,S}{d\,U} & \: = \; \frac{1}{T}
\label{label_Planck15a_eq_1} 
\; ,
\end{align}
(where $S$ is the entropy and $U$ the energy of the whole system of molecules) combined with the expression of entropy derived from the quantum 
hypothesis:{\color{blue}\footnote{$\:${\it\color{blue}Note that here Max Planck assumed the implicit hypothesis that $S=0$ for $W=1$, namely that no arbitrary additional (reference) term $S_{ref}$ is added: this is the ``\,third-law of thermodynamics\,'' defined by Max Planck since his 1900-1901 papers about the black-body radiation, which means that the entropy is equal to $0$ at very law temperature where $W=1$ for the more stable (solid) state of every bodies.  
Another way to write (\ref{label_Planck15a_eq_2}) is: 
$S - S_{ref} = k\:\ln(W/1)$ with $S_{ref} = k\:\ln(1) = 0$ / P. Marquet}}}
\vspace*{0mm}
\begin{align}
S & \: = \; k \: \ln(W)
\label{label_Planck15a_eq_2} 
\; , 
\end{align}
where $W$ means the ``\,thermodynamic\,'' probability of the stationary state (an integer 
positive number).{\color{blue}\footnote{$\:${\it\color{blue}Note that $W$ is not a true probability (namely with $0 \leq W \leq 1$). Rather, it is the ``\,number of complexions\,'' $\Omega$ introduced by Boltzmann in 1877, and replaced by the letter $W$ by Planck when he first wrote the formula (\ref{label_Planck15a_eq_2}) in the years 1900-1901 / P. Marquet}}}

A full calculation of $W$ based on the quantum hypothesis has so far only been carried out for systems of molecules with a single degree of freedom (namely for rectilinear oscillators and for rigid dipoles with a fixed axis of rotation perpendicular to the axis of symmetry).
Since such structures probably do not occur directly in nature, it will be of interest to develop a method by which the thermodynamic probability {\it\color{blue}$(W)$} can be calculated for molecules with several degrees of freedom.

It goes without saying that here the propositions valid for a single degree of freedom must be generalised in a certain way, and furthermore it is to be expected that this generalisation will not be clearly prescribed in every respect, but that it will require the introduction of certain assumptions, the justification of which can only be provided by success.
But on the other hand, it is by no means the case that one could ultimately obtain every desired result through appropriately {\it\color{blue}(arbitrary)} chosen assumptions.
On the contrary, the possibilities are extremely limited, and confirmation by experience of the formulas obtained for specific heat would be seen as a weighty testimony in favor of the theory.
Another, although not directly experimental, but nevertheless very rigorous test of the developed theorems should be briefly mentioned at the end of this {\it\color{blue}(first)} paper (§\,\ref{Section-17a-8}).


\section{\underline{-- Planck (1915a) -- §\,1 {\it\color{blue}-- Characterization of States}} (p.408-410)}
\label{Section-17a-1}
\vspace*{-2mm}

In order to characterize any state of a large number $N$ of identical molecules, each with $f$ degrees of freedom, in a statistical sense it is firstly necessary that the entire state region of a single molecule (i.e. the entire infinite $2\:f$-dimensional space formed by the coordinates 
($\varphi_1, \:\varphi_2, \:..., \:\varphi_f$) 
and the associated momenta 
($\psi_1,\:\psi_2, \:..., \:\psi_f$) of the molecules), 
is divided into certain areas (the so-called elementary areas the probability), and secondly that we know how many of the $N$ molecules in the state in question are located within the individual elementary regions.

If one denotes the elementary regions in sequence with 
$0,\:1,\:2,\:3,\:...,\:n,\:...$,$\,$\footnote{$\:$In my last work I had denoted the elementary domains $1,\:2,\:..., \:n, \:...,$ (without an elementary domain $0$), which gives some equations a slightly different look.}
then any state of the system is determined by the molecule numbers $N_0,\:N_1,\:N_2,\:...,\:N_n,\:...,$ where 
\vspace*{0mm}
\begin{align}
\sum_{n=0}^{\infty} N_n & \: = \; N
\label{label_Planck15a_eq_3} 
\; .
\end{align}
What is characteristic of the quantum hypothesis (in contrast to the classical theory) is that the elementary areas of probability have very specific shapes and sizes: their boundaries are designated by certain $(2\:f-1)$-dimensional hypersurfaces, the position of which depends solely on the nature of the molecules under consideration.
Their determination is the most important, but also the most difficult part of the task.
The simplest case is that these hypersurfaces are also the surfaces of constant energy, i.e. that all states of the molecules that correspond to one and the same energy also lie in one and the same elementary region.
We will limit ourselves to this premise in this essay. 
It is indeed realised in some cases, if certainly not in all.

Then the equations of the interfaces of the elementary regions are of the form: 
      $$ u\:=\:u_0\:=\:0\:,
 \;\;\;\;u\:=\:u_1\:,
 \;\;\;\;u\:=\:u_2\:,
 \;\;\;\; ...\,,
 \;\;\;\;u\:=\:u_n\:,
 \;\;\;\; ...\,,
 $$
where $u$ is the energy of the molecules, as a function of $\varphi$ and $\psi$, and $u_1, \:..., \:u_n, \:...$, certain constants, ordered by increasing magnitude.
Each of these areas of constant energy includes all areas with smaller ordinal numbers $n$, down to the area $u=0$, which shrinks into a single point, and the space between each two neighbouring areas forms an elementary area of probability {\it\color{blue}$(W)$}, such that the elementary area $n$ is bounded by the areas $u=u_n$ and $u=u_{n+1}$.

The amount of the constant $u_n$ is now most easily determined by the ``\,volume\,'' of the entire $2\:f$-dimensional space enclosed by the area $u=u_n$, leading to: 
\vspace*{0mm}
\begin{align}
\bigintsss  \!\!\!
\bigintsss  \!\!\!
\:.\,.\,.\: \!\!\!
\bigintsss_{\;u=0}^{\,u=u_n}\!
d\,\varphi_1 \;.\;
d\,\varphi_2 \;.\;.\;.\;
d\,\varphi_f \;\;.\;\;
d\,\psi_1 \;.\;
d\,\psi_2 \;.\;.\;.\;
d\,\psi_f \;.\;
& \: = \; \left( n \: h \right)^{f}  
\label{label_Planck15a_eq_4}
\; ,
\end{align}
where $h$ represents the universal quantum of action. 
So if we call the sizes of the elementary regions 
$G_0,\:G_1,\:...,\:G_n,\:...,$ then 
\vspace*{-2mm}
\begin{align}
G_n & \: = \; 
\left\{\;
\left(n\:+\:1\right)^{f} \;-\; \left(n\right)^{f}
\;\right\} \,.\; h^f
\label{label_Planck15a_eq_5}
\; .
\end{align}
Of course, the size and shape of the elementary areas is independent of the choice of coordinates $\varphi$.

The elementary region $0$ has the size 
\vspace*{0mm}
\begin{align}
G_0 & \: = \;  h^f
\label{label_Planck15a_eq_6}
\; ,
\end{align}
so you can also write 
\vspace*{-2mm}
\begin{align}
G_n & \: = \; 
\left\{\;
\left(n\:+\:1\right)^{f} \;-\; \left(n\right)^{f}
\;\right\} \,.\; G_0
\label{label_Planck15a_eq_7}
\; .
\end{align}
With these statements the essence of the theory to be developed here is stated.

\section{\underline{-- Planck (1915a) --  §\,2 {\it\color{blue}-- Entropy and Number of Complexions}} (p.410-411)}
\label{Section-17a-2}
\vspace*{-2mm}

The probability $W$ of a state defined by the molecular numbers $N_0,\:N_1,\:N_2,\:...,\:N_n,\:...,$ results from the consideration that the state can be realised in very different ways, depending on whether a particular molecule in question lies in this or that elementary region.

If all molecules are numbered from $1$ to $N$, and if a special distribution of the molecules over the elementary regions (in which not only the total number of molecules in an elementary region but also the numbers they carry are taken into account) is called a complexion, then a particular state comprises a large number of different complexions.

If all elementary regions are of equal size, the probability $W$ of a state is simply equal to the number of complexions corresponding to the state, namely 
\vspace*{0mm}
\begin{align}
& \frac{N\,!}{N_0\,!\;N_1\,!\;.\:.\:.\;N_n\,!\;.\:.\:.}
\label{label_Planck15a_eq_8}
\;\: .
\end{align}
However, if they are of different sizes, it must be borne in mind that when the molecules are distributed over the elementary regions, the probability of a particular molecule falling into a region with a larger $G$ is greater from the outset than the probability of it falling into a region with a smaller $G$, and, according to (\ref{label_Planck15a_eq_7}), the (thermodynamic, integer) probability of the molecule falling into the region $G_n$ is equal to 
\vspace*{0mm}
\begin{align}
\frac{G_n}{G_0} & \: = \;
\left(n\:+\:1\right)^{f} \;-\; \left(n\right)^{f}
\; = \; p_n
\label{label_Planck15a_eq_9}
\;\: .
\end{align}
Then the probability of every complexion that belongs to the state $N_0,\:N_1,\:N_2,\:...,\:N_n,\:...,$  is 
\vspace*{0mm}
\begin{align}
& p_0^{N_0}\;\;p_1^{N_1}\;\;p_2^{N_2}
\;\;.\:.\:.\;\;\;p_n^{N_n}
\;\;.\:.\:.\;\:,
\label{label_Planck15a_eq_10}
\end{align}
and thus the desired probability of the state, by multiplying (\ref{label_Planck15a_eq_8}) and (\ref{label_Planck15a_eq_10}) :
\vspace*{0mm}
\begin{align}
W & \: = \;
p_0^{N_0}\;\;p_1^{N_1}\;\;p_2^{N_2}
\;\;.\:.\:.\;\;\;p_n^{N_n}
\;\;.\:.\:.\;\;
\frac{N\,!}{N_0\,!\;N_1\,!\;.\:.\:.\;N_n\,!\;.\:.\:.}
\label{label_Planck15a_eq_11}
\;\: .
\end{align}

From this finally follows, according to (\ref{label_Planck15a_eq_2}), 
the entropy of the system in the given 
state:{\color{blue}$\,$\footnote{$\:${\it\color{blue}Note that the entropy $S$ is expressed in (\ref{label_Planck15a_eq_12}) in terms of what is nowadays called Kullack function, or Contrast, or relative entropy, namely 
$K(w,p)=\sum_n \: w_n \:.\:\ln({w_n}/{p_n})$ 
depending on the two sets of state numbers $w_n$ and $p_n$
/ P. Marquet}.}}
\vspace*{0mm}
\begin{align}
S & \: = \; 
-\:k\:N \:.\: 
\sum_{n=0}^{\infty} \: 
w_n \:.\:\ln\left(\frac{w_n}{p_n} \right)
\label{label_Planck15a_eq_12}
\;\: ,
\end{align}
whereby for abbreviation the ``\,distribution numbers\,'' 
\vspace*{0mm}
\begin{align}
w_n & \: = \; \frac{N_n}{N}
\label{label_Planck15a_eq_13}
\;\: 
\end{align}
are introduced, which according to (\ref{label_Planck15a_eq_3}) satisfy the condition: 
\begin{align}
\sum_{n=0}^{\infty} \: w_n & \: = \; 1
\label{label_Planck15a_eq_14}
\;\: .
\end{align}


\vspace*{2mm}
\section{\underline{-- Planck (1915a) -- §\,3 {\it\color{blue}-- Energy, Entropy, $\Psi$, Heat capacity}} (p.411-412)}
\label{Section-17a-3}
\vspace*{-2mm}

The further treatment of the problem is carried out according to the known methods. 
The first step is to determine the \dashuline{steady state}, i.e. the state that corresponds to the \dashuline{maximum entropy at constant total energy $U$}. 

The energy $U$ of the system is of the 
form:{\color{blue}$\,$\footnote{$\:${\it\color{blue}In fact, the energy can only be defined up the arbitrary constant (reference) term $\:U_{ref} \,=\,N \: u_{ref}$  / P. Marquet}.}}
\begin{align}
\!\!\!\!
U & \: = \; 
\sum_{n=0}^{\infty} \: N_n \;\: \overline{u}_n \; , 
{\color{blue}
         \left[\,
         \mbox{or:}\;\;
U \,-\, U_{ref} \, = \, 
\sum_{n=0}^{\infty} \: N_n \: 
\left( \overline{u}_n - u_{ref} \right)
\;\mbox{with}\;\;
U_{ref} \,=\, 
\sum_{n=0}^{\infty} \: N_n \; u_{ref}
\,=\, N \: u_{ref} 
         \,\right]
}\!\!
\label{label_Planck15a_eq_15}
\end{align}
where $\overline{u}_n$ (not to be confused with $u_n$) denotes the average energy of all molecules located in the elementary regions $n$.
This quantity has a very specific value that is independent of the number of molecules, since within each elementary region the distribution density of the molecules is uniform.
Therefore: 
\vspace*{-4mm}
\begin{align}
\overline{u}_n & \: = \; 
\frac{\displaystyle 
\bigintsss_{\;u_n}^{\,u_{n+1}}
\!\!\!\! u \; d\,G}{G_n}
\label{label_Planck15a_eq_16}
\; ,
\end{align} 
with the use of the abbreviation: 
\vspace*{2mm}
\begin{align}
d\,G & \: = \; 
\bigintsss  \!\!\!
\bigintsss  \!\!\!
\:.\,.\,.\: \!\!\!
\bigintsss_{\;u}^{\,u+du}\!
d\,\varphi_1 \;.\;
d\,\varphi_2 \;.\;.\;.\;
d\,\varphi_f \;\;.\;\;
d\,\psi_1 \;.\;
d\,\psi_2 \;.\;.\;.\;
d\,\psi_f
\label{label_Planck15a_eq_17}
\; .
\end{align}

According to (\ref{label_Planck15a_eq_14}) and  (\ref{label_Planck15a_eq_1}), the \dashuline{condition of the maximum of entropy} {\it\color{blue}$S$ given by (\ref{label_Planck15a_eq_12})} \dashuline{at constant $U$} {\it\color{blue}given by (\ref{label_Planck15a_eq_15})} gives the distribution numbers in the 
\dashuline{stationary state}:{\color{blue}$\,$\footnote{$\:${\it\color{blue}See below the demonstration of this relationship (\ref{label_Planck15a_eq_18})  / P. Marquet}.}}
\vspace*{-1mm}
\begin{align}
\boxed{\;
w_n \; = \; 
\frac{\displaystyle 
p_n \:.\: \exp\left( 
-\,\frac{\overline{u}_n}{k\:T} \right)
}
{\displaystyle  
\sum_{n=0}^{\infty} \;
p_n \:.\: \exp\left( 
-\,\frac{\overline{u}_n}{k\:T} \right)
}
\;}
\label{label_Planck15a_eq_18}
\; ,
\end{align}
and then \dashuline{the energy of the system} as a function of temperature: 
\vspace*{0mm}
\begin{align}
U \:=\; N \; \overline{u}
\:=\; N \; 
\sum_{n=0}^{\infty} \; w_n \; \overline{u}_n
 & \: = \; 
 N \:.\:
\frac{\displaystyle 
\sum_{n=0}^{\infty} \;
p_n \; \overline{u}_n \; \exp\left( 
-\,\frac{\overline{u}_n}{k\:T} \right)
}
{\displaystyle  
\sum_{n=0}^{\infty} \;
p_n \; \exp\left( 
-\,\frac{\overline{u}_n}{k\:T} \right)
}
\label{label_Planck15a_eq_19}
\; ,
\end{align}
and then, according to (\ref{label_Planck15a_eq_12}), \dashuline{the entropy {\it\color{blue}(of the system)}}:
\vspace*{0mm}
\begin{align}\tag{{\color{blue}19bis}}
S \:=\; N \; \overline{s} 
  \:=\; N \:.\: \left\{\;
  \frac{\overline{u}}{T}
  \;+\;
  k \: \ln\left[\;
\sum_{n=0}^{\infty} \;
p_n \; \exp\left( 
-\,\frac{\overline{u}_n}{k\:T} \right)
\;\right]
\;\right\}
\label{label_Planck15a_eq_19bis}
\; .
\end{align}

{\it\color{blue}
\vspace*{-4mm}
\begin{center}
{\bf---------------------------------------------------}
\end{center}
\vspace*{-3mm}

\hspace*{70mm}(Notes of P. Marquet)

Max Planck did not give the demonstration for the special relationship (\ref{label_Planck15a_eq_18}) for 
$w_n=N_n/N$ given in terms of 
$p_n = \left(n\:+\:1\right)^{f} -
       \left(n\right)^{f}$ 
and $\overline{u}_n$.
The aim of these additional notes is to provide a demonstration of this relationship (\ref{label_Planck15a_eq_18}), a demonstration that can be obtained obtained/checked from the following considerations. 

- The constraint (\ref{label_Planck15a_eq_14}) corresponds to
$\:\sum_{n=0}^{\infty} \: K \: dw_n = 0$, whatever the arbitrary constant $K$ may be.
   
- From (\ref{label_Planck15a_eq_15}) the energy can be rewritten as $\;U(w_n,\overline{u}_n) = N \; \sum_{n=0}^{\infty} \: w_n \: \overline{u}_n$, due to (\ref{label_Planck15a_eq_13}) and $N_n = N \: w_n$.

- From (\ref{label_Planck15a_eq_12}) the entropy $\;S(w_n,p_n)$ is written in terms of $w_n$ and $p_n$.

- The search for the maximum of $\;S(w_n,p_n)$ at constant 
$\;U(w_n,\overline{u}_n)$ must be understood and fulfilled at constant $T$ (or $k\:T$), and thus with (\ref{label_Planck15a_eq_1}) possibly rewritten as:
\vspace*{0mm}
$$ d\left[\: \frac{S(w_n,p_n)}{k} \:\right] \;=\; 
 d\left[\: \frac{U(w_n,\overline{u}_n)}{k\:T} \:\right] 
   \;=\; 0 \; . $$
\vspace*{-6mm}

- Accordingly, the change in energy (at constant $k\:T$) is:
\vspace*{0mm}
\begin{align}
d\left[\: \frac{U(w_n,\overline{u}_n)}{k\:T} \:\right] 
& \:=\; 
N \:.\: \left[\;
\sum_{n=0}^{\infty} \: 
    w_n \; 
    d\left(\frac{\overline{u}_n}{k\:T}\right)
\;+\;
\sum_{n=0}^{\infty} \: 
    \left(\frac{\overline{u}_n}{k\:T}\right) \; 
    d\left(w_n\right)
\;\right]
\nonumber
\; , \\
d\left[\: \frac{U(w_n,\overline{u}_n)}{k\:T} \:\right] 
& \:=\; 
N \:.\: \sum_{n=0}^{\infty} \:\left[\;
    w_n \; 
    \left(\frac{d\,\overline{u}_n}{k\:T}\right)
\;+\;
    \left(\frac{\overline{u}_n}{k\:T}\right) \; 
    d\left(w_n\right)
\;\right]
\nonumber
\; , \\
d\left[\: \frac{U(w_n,p_n)}{k\:T} \:\right] 
& \:=\; 
N \:.\: \sum_{n=0}^{\infty} \:\left[\;
    w_n \; 
  \left(
    \left.\frac{\partial\,\overline{u}_n}
               {\partial\,p_n}\right|_{w_n}
    \frac{dp_n}{k\:T}
    \:+\:
    \left.\frac{\partial\,\overline{u}_n}
               {\partial\,w_n}\right|_{p_n}
    \frac{dw_n}{k\:T}
  \right)
\;+\;
    \left(\frac{\overline{u}_n}{k\:T}\,+\,K'\right) \; 
    d\left(w_n\right)
\;\right]
\nonumber
\; ,\\
d\left[\: \frac{U(w_n,p_n)}{k\:T} \:\right] 
& \:=\; 
N \:.\: \sum_{n=0}^{\infty} \:\left[\;\:
  \boxed{
    \frac{w_n\:p_n}{k\:T} \:.\:
    \left.\frac{\partial\,\overline{u}_n}
               {\partial\,p_n}\right|_{w_n}
  }\;\;
    d\ln(p_n)  
\;+\;
   \dbox{\ensuremath{\displaystyle\:
      \frac{\overline{u}_n}{k\:T}
      \,+\,K'
      \,+\, \frac{w_n}{k\:T}\:.\:
    \left.\frac{\partial\,\overline{u}_n}
               {\partial\,w_n}\right|_{p_n}
    }} \;\;
    d\left(w_n\right)
\;\right]
\nonumber
\; ,
\end{align}
where $K'$ is an arbitrary constant introduced due to $\:\sum_{n=0}^{\infty} \: K' \: dw_n = 0$.

- Similarly, the change in entropy is:
\vspace*{0mm}
\begin{align}
d\left[\: \frac{S(w_n,p_n)}{k} \:\right] 
& \:=\; 
N \:.\: \left[\;
\sum_{n=0}^{\infty} \: 
    w_n \; 
    d\ln(p_n)
\;-\;
\sum_{n=0}^{\infty} \: 
    \ln\!\left(\frac{w_n}{p_n}\right) \; 
    d\left(w_n\right)
\;-\;
\sum_{n=0}^{\infty} \: 
    d\left(w_n\right)
\;\right]
\nonumber
\; , \\
d\left[\: \frac{S(w_n,p_n)}{k} \:\right] 
& \:=\; 
N \:.\: \left[\;
\sum_{n=0}^{\infty} \: 
    w_n \; 
    d\ln(p_n)
\;-\;
\sum_{n=0}^{\infty} \: 
    \ln\!\left(\frac{w_n}{p_n}\right) \; 
    d\left(w_n\right)
\;-\;
\sum_{n=0}^{\infty} \: 
    K \: d\left(w_n\right)
\;\right]
\nonumber
\; , \\
d\left[\: \frac{S(w_n,p_n)}{k} \:\right] 
& \:=\; 
N \:.\: \sum_{n=0}^{\infty} \: \left[\;\:
  \boxed{\; \phantom{\frac{A}{B}}\!\!\!\!\!\!
    w_n \; 
  }\;
    d\ln(p_n)
\;+\;
   \dbox{\ensuremath{\displaystyle\;
    \ln\!\left(\frac{p_n}{w_n \: K}\right)\: 
   }}\;\;
    d\left(w_n\right)
\;\right]
\nonumber
\; ,
\end{align}
where $K$ is an arbitrary constant introduced due to $\:\sum_{n=0}^{\infty} \: K \: dw_n = 0$.

- In order to ensure the relationship (\ref{label_Planck15a_eq_1}) at constant $k\:T$, and thus $\:d\,U/(k\:T)=d\,S/k$, it is possible to assumed that the $n$-term-by-term equalities are valid by equating the solid and dashed boxes, leading to:
\vspace*{0mm}
\begin{align}
  \boxed{
    \frac{w_n\:p_n}{k\:T} \:.\:
    \left.\frac{\partial\,\overline{u}_n}
               {\partial\,p_n}\right|_{w_n}
  }\;\;
 & \:=\;\:
  \boxed{\; \phantom{\frac{A}{B}}\!\!\!\!\!\!
    w_n \; 
  }\;
\nonumber
\; , \\
\mbox{and\;\;\;\;}
   \dbox{\ensuremath{\displaystyle\:
      \frac{\overline{u}_n}{k\:T}
      \,+\,K'
      \,+\, \frac{w_n}{k\:T}\:.\:
    \left.\frac{\partial\,\overline{u}_n}
               {\partial\,w_n}\right|_{p_n}
    }} \;\;
 & \:=\;\:
   \dbox{\ensuremath{\displaystyle\;
    \ln\!\left(\frac{p_n}{K \; w_n}\right)\: 
   }}\;\;
\nonumber
\; .
\end{align}

- It is easy to checked that $\overline{u}_n$ given by (\ref{label_Planck15a_eq_18}), and thus 
\vspace*{-1mm}
\begin{align}
\overline{u}_n(w_n,\:p_n) & \: = \; k\:T \;.\;
\ln\!\left(\;
  \frac{p_n}{K\;w_n}
\;\right)
   \nonumber 
\; ,
\end{align}
with $K$ a constant, is compatible with the previous two equalities between the solid and dashed boxes, 
where $K'=1$ is indeed a true constant and where the second constant 
$$\:K \; = \; \sum_{n=0}^{\infty} \; p_n \:.\: 
\exp\left( -\,\frac{\overline{u}_n}{k\:T} \right)$$ 
corresponds to the denominator of (\ref{label_Planck15a_eq_18}), and is simply determined by the relationship (\ref{label_Planck15a_eq_14}) and thus by the normalisation formula $\:\sum_{n=0}^{\infty} \: w_n = 1$. 

- Note that $K$ can indeed be considered as a constant from (\ref{label_Planck15a_eq_19bis}), which can be rewritten  as 
\begin{align}\tag{19ter}
\overline{s} \;=\; 
   \frac{\overline{u}}{T} \;+\; k\:\ln(K) 
 \label{label_Planck15a_eq_19ter}
\; , 
\end{align}
and from the hypothesis of constant energy at constant $T$, namely $\:d\,(\overline{u}\,/\,T)=0$, and also from the hypothesis of a maximum for $\:\overline{s}$, namely at least with $\:d\,\overline{s}=0$, which jointly lead from (\ref{label_Planck15a_eq_19ter}) to $\,d\ln(K)=(d\,K)/K=0\:$, and thus to $\:d\,K=0\:$.

- The next partial derivatives will be computed for the normalized entropy $\:s_{\ast}(w_n,p_n)$ defined from the entropy  $S$ given by (\ref{label_Planck15a_eq_12}) leading to the relative-entropy (Kullback) function: 
\vspace*{-1mm}
\begin{align}
s_{\ast}(w_n,p_n) 
\;=\; \frac{S(w_n,p_n)}{k\:N}
\;=\; \sum_{n=0}^{\infty} \: 
w_n \:.\:\ln\left(\frac{p_n}{w_n} \right)
\nonumber 
\; .
\end{align}
\vspace*{-5mm}

- The last property to be proven is the fact that both $S(w_n,p_n)$ and $s_{\ast}(w_n,p_n)$ are maximum, namely that not only $\:d\,s_{\ast}(w_n,p_n)=0$ (see its impact in the previous item for proving $\:d\,K=0\,$), but also to check that $d^{\,2}\,s_{\ast}(w_n,p_n)<0$ (still at constant $T$), according to the left part of the Fig.~\ref{fig_Planck_1915a}.
\begin{figure}[hbt]
\centering
{\color{blue}
--------------------------------------------------------------------------------------------------------------
}
\vspace{1mm}
\\
\includegraphics[width=0.95\linewidth]{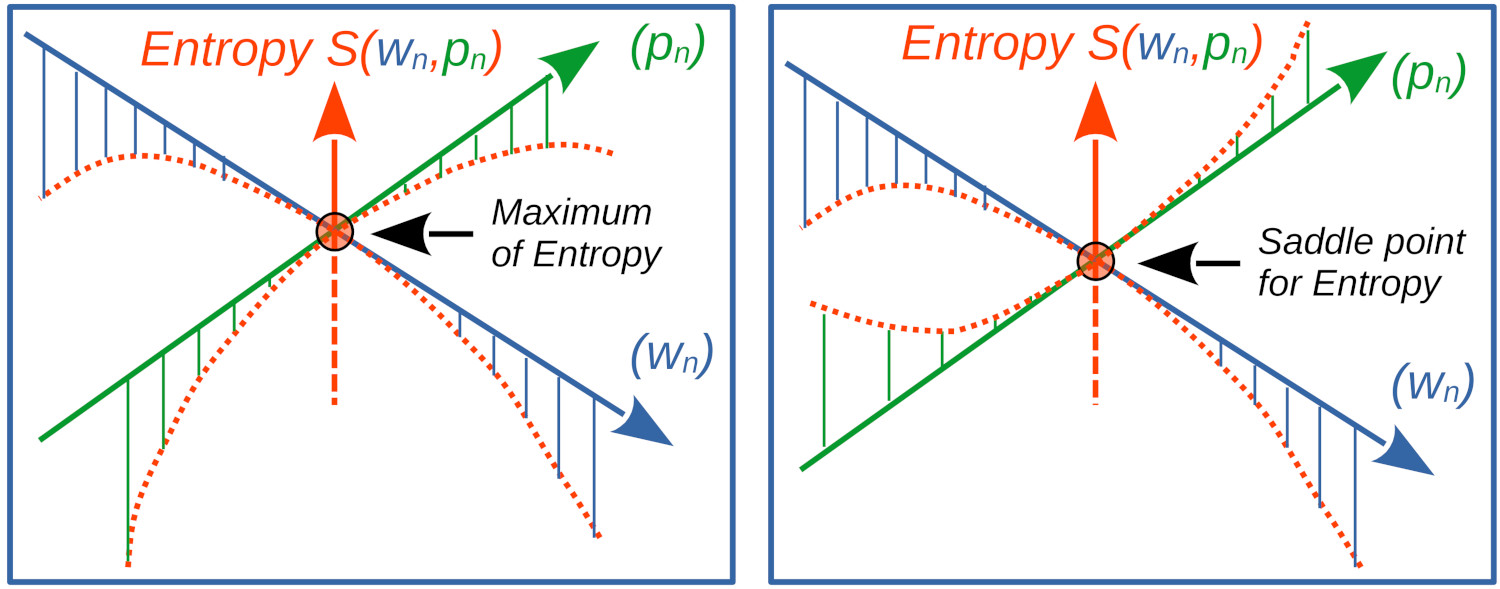} 
\vspace{-3mm}
\caption{\color{blue}\it A figure to illustrate the case of maximum of entropy $S$ (on the left) or the Saddle-point feature (on the right), with in both cases $dS=0$ locally around the origin of the axes $(w_n,\,p_n)$ / P. Marquet
\label{fig_Planck_1915a}}
{\color{blue}
--------------------------------------------------------------------------------------------------------------
}
\end{figure}
\vspace*{-2mm}

\noindent
Sufficient criteria for $s_{\ast}(w_n,p_n)$ to be a 
maximum$\,${\color{blue}\footnote{\color{blue}$\:$See: 
\url{https://en.wikipedia.org/wiki/Second_partial_derivative_test}}}
are:
\vspace*{-2mm}
\begin{align}
Det(H) \;=\; D(w_n,p_n) 
\;=\;
\left(
\frac{\partial^{\,2}\,s_{\ast}}{\partial\,w_n\;\partial\,w_n}
\;\;.\;\;
\frac{\partial^{\,2}\,s_{\ast}}{\partial\,p_n\;\partial\,p_n}
\right)
\;-\;
\left(
  \frac{\partial^{\,2}\,s_{\ast}}{\partial\,w_n\;\partial\,p_n}
\right)^{2}
\;>\; 0
 \nonumber 
\; , \hspace*{10mm} \\
\mbox{and} \quad \quad
\frac{\partial^{\,2}\,s_{\ast}}{\partial\,w_n\;\partial\,w_n}
\;<\; 0
\quad\mbox{or}\quad
\frac{\partial^{\,2}\,s_{\ast}}{\partial\,p_n\;\partial\,p_n}
\;<\; 0
\quad\mbox{or}\quad
tr(H) \;=\;
\frac{\partial^{\,2}\,s_{\ast}}{\partial\,w_n\;\partial\,w_n}
\;+\;
\frac{\partial^{\,2}\,s_{\ast}}{\partial\,p_n\;\partial\,p_n}
\;<\; 0
 \nonumber 
\; ,
\end{align}
where $Det(H)$ is the determinant 
and $tr(H)$ is the trace of the Hessian matrix 
$H_{\,i\,j} = 
{\partial^{\,2}\,s_{\ast}}/{(\partial\,x_i\:\partial\,x_j)}$,
with $x_1=w_n$ and $x_2=p_n$.
\vspace*{1mm}

The first- and second-order partial derivatives of the relative-entropy (Kullback) function are:
\vspace*{-2mm}
\begin{align}
\left.\frac{\partial\,s_{\ast}}{\partial\,p_n}\right|_{w_n}
=\; \sum_{n=0}^{\infty} \: \frac{w_n}{p_n} 
\;,\quad
\dbox{\ensuremath{\displaystyle
\left.
\frac{\partial^{\,2}\,s_{\ast}}{\partial\,w_n\;\partial\,p_n}
\right|_{w_n,\:p_n}
=\; \sum_{n=0}^{\infty} \: \frac{1}{p_n}
}}
\;,\quad
\boxed{
\left.
\frac{\partial^{\,2}\,s_{\ast}}{\partial\,p_n\;\partial\,p_n}
\right|_{w_n,\:w_n}
=\; \sum_{n=0}^{\infty} \: \left(-\,\frac{w_n}{p_n^2} \right)
}
\nonumber \; , \\
\left.\frac{\partial\,s_{\ast}}{\partial\,w_n}\right|_{p_n}
=\; \sum_{n=0}^{\infty} \: 
 \left[\: \ln\left(\frac{p_n}{w_n}\right) - 1 \:\right] 
\;,\quad
\dbox{\ensuremath{\displaystyle
\left.
\frac{\partial^{\,2}\,s_{\ast}}{\partial\,p_n\;\partial\,w_n}
\right|_{p_n,\:w_n}
=\; \sum_{n=0}^{\infty} \: \frac{1}{p_n}
}}
\;,\quad
\boxed{
\left.
\frac{\partial^{\,2}\,s_{\ast}}{\partial\,w_n\;\partial\,w_n}
\right|_{p_n,\:p_n}
=\; \sum_{n=0}^{\infty} \: \left(-\,\frac{1}{w_n} \right)
}
\nonumber \; .
\end{align}

The two synonymous Cauchy-Schwarz 
inequalities$\,${\color{blue}\footnote{\color{blue}$\:$See: \url{https://en.wikipedia.org/wiki/Cauchy\%E2\%80\%93Schwarz_inequality}}}
\vspace*{0mm}
\begin{align}
  \frac{1}{2} \: 
 \sum_{i=0}^{N\rightarrow\infty} 
 \sum_{j=0}^{N\rightarrow\infty}
 \left(\: u_i\;v_j \;-\; u_j\;v_i \:\right)^2 
 \; \geq \; 0 \;  \hspace*{10mm}  &
 \nonumber \\
\mbox{and thus}\quad
 \left(\; 
 \sum_{n=0}^{N\rightarrow\infty} 
  \:u_n^2
 \,\right)
 \;.\;
 \left(\; 
 \sum_{n=0}^{N\rightarrow\infty} 
  \:v_n^2
 \,\right)
    \;-\;
 \left(\; 
 \sum_{n=0}^{N\rightarrow\infty} 
  \:u_n\;v_n
 \,\right)^2 
& \: \geq \; 0 
 \nonumber 
\;  
\end{align}
can be applied to 
$u_n=\sqrt{\,w_n}/p_n$ and 
$v_n=1/\sqrt{\,w_n}$
(with $w_n=N_n/N>0$),
leading to:
\vspace*{-1mm}
\begin{align}
  \left(\; 
 \sum_{n=0}^{N\rightarrow\infty} 
  \: \frac{-\,w_n}{p_n^2}
 \,\right)
 \;.\;
 \left(\; 
 \sum_{n=0}^{N\rightarrow\infty} 
  \: \frac{-\,1}{w_n}
 \,\right)
    \;-\;
\;
 \left(\; 
 \sum_{n=0}^{N\rightarrow\infty} 
  \: \frac{1}{p_n}
 \,\right)^2
& \:=\;
   Det(H) \;=\; D(w_n,p_n)
 \; \geq \; 0
 \nonumber 
\; ,
\end{align}
where the two minus signs (with $-1 \times -1 = 1$) have been introduced (without lack of generality) in the first term (the product of the sums), in order to correspond to the second order partial derivatives of $s_{\ast}$ (the solid boxes).
 
Written as such, the previous inequality corresponds to 
$D(w_n,p_n) \geq 0$ 
valid for the relative-entropy (Kullback) variable 
$\:s_{\ast}\,(w_n,p_n)$, 
and thus for the entropy variables 
$\:\overline{s}\,(w_n,p_n)=k\:s_{\ast}\,(w_n,p_n)$ and 
$\:S\,(w_n,p_n)=k\:N\:s_{\ast}\,(w_n,p_n)$.

Therefore, since the (solid boxes) second order partial derivatives of $s_{\ast}$ are negative (due to both ${-\,w_n}/{p_n^2}<0$ and ${-\,1}/{w_n}<0$), the status of a maximum entropy (see the left part of the Fig.~\ref{fig_Planck_1915a}) is valid for all the cases where $Det(H) = D(w_n,p_n) > 0$.

Only for the case where $Det(H) = D(w_n,p_n) = 0$ the entropy $s_{\ast}(w_n,\,p_n)$ might be a saddle point (see the right part of the Fig.\,\ref{fig_Planck_1915a}). 
The last question is thus to know whether this case $Det(H) = D(w_n,p_n) = 0$ may exit, or not.
To do so, the first Cauchy-Schwarz inequality can be used
together with $u_n=\sqrt{\,w_n}/p_n$ and 
$v_n=1/\sqrt{\,w_n}$ to study the condition for an equality: 
\vspace*{-1mm}
\begin{align}
 \sum_{i=0}^{N\rightarrow\infty} 
 \sum_{j=0}^{N\rightarrow\infty}
 \left(\: u_i\;v_j \;-\; u_j\;v_i \:\right)^2 
 \; = \; 
 \sum_{i=0}^{N\rightarrow\infty} 
 \sum_{j=0}^{N\rightarrow\infty}
 \left(\: 
   \frac{\sqrt{\,w_i}}{p_i}\;
   \frac{1}{\sqrt{\,w_j}} 
 \;-\; 
   \frac{\sqrt{\,w_j}}{p_j}\;
   \frac{1}{\sqrt{\,w_i}} 
 \:\right)^2 
 \; = \; 0 \; 
 \nonumber 
\; ,
\end{align}
which corresponds to:
\vspace*{-3mm}
\begin{align}
 \sum_{i=0}^{N\rightarrow\infty} 
 \sum_{j=0}^{N\rightarrow\infty}
 \:\frac{\left(\:
  p_j\:w_i\;-\;p_i\:w_j
 \:\right)^2}
 {p_i^2\;w_j\;p_j^2\;w_i}
 \; = \; 0 \; 
 \nonumber 
\; .
\end{align}
Since $w_n=N_n/N>0$, the conditions for this equality is that all numerators would cancel out, namely that $p_j\:w_i\;=\;p_i\:w_j$ for all $\:i, \:j$ from $0$ to $N \rightarrow \infty$, or equivalently from (\ref{label_Planck15a_eq_18}):
\vspace*{0mm}
\begin{align}
 \frac{w_i}{p_j} 
 \; = \; 
 \frac{1}{K} \;
 \exp\left( 
   -\,\frac{\overline{u}_i}{k\:T} 
 \right)
 \; = \; 
 \frac{w_j}{p_j} 
 \; = \; 
 \frac{1}{K} \;
 \exp\left( 
   -\,\frac{\overline{u}_j}{k\:T} 
 \right)
\quad
\forall (i,j) \in [\:0, \, N\rightarrow\infty \:]^2
 \nonumber 
\; .
\end{align}
Since both $T$ and $K$ are constant (see above), this means that the saddle-point case for the entropy corresponds to the special case for which all the energy states $\overline{u}_n = U_0/N$ are identical $\:\forall\,n$. 

However, since for a constant energy $U_0$ and for infinitely large $N$ the energy for individual states $\overline{u}_n$ becomes infinitely small, this distribution of energy over the states seems unrealistic ($\,U_0\approx 0 \times \infty\,$) and without physical meaning. 
Moreover, for this case and from (\ref{label_Planck15a_eq_12}) and (\ref{label_Planck15a_eq_18}), the normalized entropy is:
\vspace*{0mm}
\begin{align}
S  
& \,=\, -\:N\:k\:\sum_{n=0}^{\infty} \: w_n \:.\:
         \ln\left(\frac{w_n}{p_n} \right)
\,=\, -\:N\:k\:\sum_{n=0}^{\infty} \: w_n \:.
 \left(
   -\,\frac{U_0}{N\:k\:T} \,-\, \ln(K) 
 \right)
\nonumber \: , \\
S  
& \,=\, 
  \underbrace{
      \left(\sum_{n=0}^{\infty} 
      \: w_n\right) 
  }_{\displaystyle (=1)}
    \:.\: \left[\: 
          \frac{U_0}{T} \,+\,N\:k\:\ln(K) 
          \:\right]
\:=\; \frac{U_0}{T} \,+\, N\:k\:\ln(K) 
\nonumber \: .
\end{align}
\vspace*{-6mm}

\noindent
which is similar to $N$ times (\ref{label_Planck15a_eq_19ter}), but with $U/N$ replaced by $U_0/N$ and with $K$ to be computed as  
\vspace*{0mm}
\begin{align}
K & \,=\, 
\sum_{n=0}^{\infty} \; p_n \:.\: 
\exp\left( -\,\frac{\overline{u}_n}{k\:T} \right)
\;=\;
\exp\left( -\,\frac{U_0}{N\:k\:T} \right)
\:.\:
\left(\: \sum_{n=0}^{\infty} \; p_n \:\right)
\nonumber \: .
\end{align}
However, from (\ref{label_Planck15a_eq_9}) 
$\;p_n = \left(n\:+\:1\right)^{f} \,-\, \left(n\right)^{f}$ $\,[\,\propto f\:\left(n\right)^{f-1}$ for large $n\,]$ is increasing with $n$ for $f > 1$, and therefore the sum $\:\sum_{n=0}^{\infty}\;p_n\:$ tends toward $+\,\infty$ (even for $f=1$ and $p_n=1$). 
As for the exponential term, since $U_0$ is finite, $-\,U_0/N$ becomes infinitely small for large $N$ and the exponential term tends toward unity.
This shows that $\:\ln(K)\:$ and the entropy become infinite, and that the possible saddle-point case for the entropy (with constant values of $\overline{u}_n = U_0/N$) is unrealistic and has no physical meaning.

\vspace*{-4mm}
\begin{center}
{\bf---------------------------------------------------}
\end{center}
\vspace*{-4mm}

We cannot know the true method used by Max Planck to establish the formula (\ref{label_Planck15a_eq_18}). 
It may be close or similar to that proposed in this additional note, but the fact that Max Planck did not wish to describe these calculations in detail shows that he may have considered them too obvious, and he may have been too familiar with the methods of mathematical 
physics$\,${\color{blue}\footnote{$\:$\it\color{blue}Max Planck explained in a paper written in $1943$ (\,$4$ years before his death) about ``\,The history of the discovery of the physical quantum of action\,'' (see \url{https://zenodo.org/records/10513021} or \url{https://arxiv.org/abs/2403.09689}) that: 
``\,The impression his 1879 Doctoral dissertation paper (about the entropy and the second law of Clausius) had on the physical public at the time was zero. As I know from conversations with my university teachers, none of them had any understanding of its content. They probably only let it pass as a dissertation because they knew me from my other work in the \dashuline{practical course in physics and in the mathematics seminar}.\,''}}.
Anyhow, Max Planck might also has used another (variational? Lagrange multipliers?) method for the establishment of this relationship. 


\vspace*{-5mm}
\begin{center}
{\bf---------------------------------------------------}
\end{center}
\vspace*{-4mm}

Incidentally, note that the reverse of the relative-entropy (Kullback) Planck's variables 
$1/u_n=p_n/\sqrt{\,w_n}$ and 
$1/v_n=\sqrt{\,w_n}$
are similar to those
$u_n=u'_n/\sqrt{\,v'_n}$ and 
$v_n=\sqrt{\,v'_n}$
used to defined what are called Bergstr\"om's (1949) inequality, Sedrakyan's (1997) inequality, Engel's (1998) form, T2 lemma, or Titu's (Andreescu, 2003) lemma, for which: 
\vspace*{-2mm}
\begin{align}
  \left(\; 
 \sum_{n=1}^{N} 
  \: \frac{{u'}_n^{\,2}}{v'_n}
 \,\right)
 \;.\;
 \left(\; 
 \sum_{n=1}^{N} 
  \: v'_n
 \,\right)
    \;-\;
\;
 \left(\; 
 \sum_{n=1}^{N} 
  \: u'_n
 \,\right)^2
 \; \geq \; 0 \; ,
 \hspace*{10mm}
 \nonumber \\
\mbox{used as:}\quad\quad
 \frac{\left(\:u'_1 \:+\: u'_2 \:+\: u'_3 
       \:+\: .\:.\:. \:+\: u'_N \:\right)^2}
      {v'_1 \:+\: v'_2 \:+\: v'_3 
       \:+\: .\:.\:. \:+\: v'_N}
\; \leq \;
\frac{{u'}_1^{\,2}}{v'_1} \:+\: 
\frac{{u'}_2^{\,2}}{v'_2} \:+\: 
\frac{{u'}_3^{\,2}}{v'_3} \:+\: 
.\:.\:.  \:+\: 
\frac{{u'}_N^{\,2}}{v'_N}
 \nonumber 
\; 
\end{align}
to easily show, for example$\,${\color{blue}\footnote{\color{blue}$\:$See: \url{https://en.wikipedia.org/wiki/Titu\%27s_lemma}}}, that 
$(a^{4}+b^{4})\geq (a+b)^{4}/8$
or  
$\:2/(a+b)+2/(b+c)+2/(a+c) \geq 9/(a+b+c)$
or 
$\:a/(b+c)+b/(a+c)+c/(a+b) \geq 3/2$
for any positive real numbers $a$, $b$, $c$.
\vspace*{1mm}

The relative-entropy (Kullback) Planck's version of the Titu's (Andreescu, 2003) lemma is therefore:
\vspace*{0mm}
\begin{align}
 \frac{\left(\:1/p_1 \,\;+\; 1/p_2 \,\;+\; 1/p_3 
      \,\;+\; .\:.\:. \,\;+\; 1/p_N \right)^2}
 {1/{\,w_1} \:+\: 1/{\,w_2} \:+\: 1/{\,w_3} 
      \:+\: .\:.\:. \:+\: 1/{\,w_N} }
\; \leq \;
\frac{{w}_1}{{\,p_1^{\,2}} } \:+\: 
\frac{{w}_2}{{\,p_2^{\,2}} } \:+\: 
\frac{{w}_3}{{\,p_3^{\,2}} } \:+\: 
 .\:.\:. \:+\: 
\frac{{w}_N}{{\,p_N^{\,2}} }
 \nonumber \; , \hspace*{15mm} \\
 \left(\:
 \frac{1}{p_1} + \frac{1}{p_2} + \frac{1}{p_3} 
 + .\:.\:. + \frac{1}{p_N} 
 \right)^2
\; \leq \;
 \left(\:
 \frac{1}{w_1} + \frac{1}{w_2} + \frac{1}{w_3} 
 + .\:.\:. + \frac{1}{w_N}
 \right)
 \left(\:
\frac{{w}_1}{{\,p_1^{\,2}}} + 
\frac{{w}_2}{{\,p_2^{\,2}}} + 
\frac{{w}_3}{{\,p_3^{\,2}}} 
+ .\:.\:. +
\frac{{w}_N}{{\,p_N^{\,2}}} 
 \right)
 \nonumber \; .
\end{align}


\vspace*{-4mm}
\begin{center}
{\bf---------------------------------------------------}
\end{center}
\vspace*{-3mm}
}

{\color{blue}\it Continuation of the Max Planck text:}

When using the {\it\color{blue}(absolute)} temperature $T$ as an independent variable, it is often more convenient to introduce the free energy $F$ or, even better, the {\it\color{blue}(Massieu's)} thermodynamic function {\it\color{blue}(see the footnote$\,{}{\mbox{\ref{label_footnote_Masieu}}}$)}: 
\vspace*{-4mm}
\begin{align}
\Psi \:=\; -\:\frac{F}{T}
\;=\; S \:-\: \frac{U}{T}
\:=\; N \; k \; 
\ln\!\left[\;
\sum_{n=0}^{\infty} \;
p_n \; \exp\left( 
-\,\frac{\overline{u}_n}{k\:T} \right)
\;\right]
\;
{\color{blue}
\;=\; N \; k \; \ln\!\left(\: Z \:\right)
}
\label{label_Planck15a_eq_20}
\; ,
\end{align}
{\it\color{blue}where $\:Z$ is called ``\,partition function\,'' in modern statistical physics\,}.  
Then the energy is: 
\vspace*{0mm}
\begin{align}
U \;=\; -\:\frac{\partial\,\Psi}{\partial\,\tau}
{\color{blue} \:\;=\; 
T^2 \: \frac{\partial\,\Psi}{\partial\,T} }
\label{label_Planck15a_eq_21}
\; ,
\end{align}
where 
\vspace*{-2mm}
$$\tau\;=\;\frac{1}{T} \; ,$$ 
and the heat capacity of the whole system is: 
\vspace*{-2mm}
\begin{align}
\hspace*{15mm}
C \;=\; \tau^2 \; \frac{\partial^2\,\Psi}{\partial\,\tau^2}
\label{label_Planck15a_eq_22}
\; .
\end{align}



\section{\underline{-- Planck (1915a) -- §\,4 {\it\color{blue}-- Low and high temperatures}} (p.412-413)}
\label{Section-17a-4}
\vspace*{-2mm}

\dashuline{At low temperatures}, only a small number of terms need to be taken into account in the sum $\sum$.

\noindent
\dashuline{At absolute zero temperature} the entire sum is limited to the first term, and one obtains from {\it\color{blue}($\:w_0=1$ and)} (\ref{label_Planck15a_eq_19}): 
\vspace*{-1mm}
       $$ U \;=\; N \; \overline{u}_0 
       {\it\color{blue}\:\;=\; U_0} \; , $$ 
i.e. the molecules are all located in the elementary region $0$, which they fill with uniform density, and their energy, {\bf\dashuline{the zero-point energy}}, corresponds to the average energy ${\it\color{blue} U_0}$ of this region.

\dashuline{At high temperatures}, however, the number of terms to be taken into account in the sum $\sum$ is so large that the terms with medium-sized atomic numbers can be omitted entirely, and only those with large atomic numbers $n$ remain.
Then the sum can be written as an integral, since the exponential function changes so little from term to term that in the expression (\ref{label_Planck15a_eq_9}) of $p_n$ the size $G_n$ of the elementary domain $n$, although it is large compared to $G_0$, is nevertheless equal to the differential $d\,G$ of the state space, and accordingly the mean energy $\overline{u}_n$ can be set equal to the energy $u$ at any point in the domain, i.e. according to (\ref{label_Planck15a_eq_19}): 
\vspace*{-2mm}
\begin{align}
U & \: = \; 
 N \;.\:\;
\frac{\displaystyle 
\bigintsss
u \; \exp\left( 
-\,\frac{u}{k\:T} \right)
\: d\,G
}
{\displaystyle  
\bigintsss
\exp\left( 
-\,\frac{u}{k\:T} \right)
\: d\,G
}
\label{label_Planck15a_eq_23}
\;\; .
\end{align}
This is the well-known expression of classical theory, which leads to the theorem of uniform energy distribution.



\section{\underline{-- Planck (1915a) -- §\,5 {\it\color{blue}-- Rectilinear Periodic Oscillator}} (p.413-414)}
\label{Section-17a-5}
\vspace*{-2mm}

We will now briefly discuss some simple applications.

For \dashuline{a rectilinear, simply periodic oscillator} $f=1$, so according to 
(\ref{label_Planck15a_eq_5}) and 
(\ref{label_Planck15a_eq_9}):
   $$ G_n \;=\; h \: , \;\;\; p_n\;=\;1 \; , $$ 
i.e. all elementary areas are the same size, and their size is equal to the elementary constant of action. 
Furthermore: 
   $$ u \; =\; \frac{m}{2} \: 
     \left( \: \dot{\varphi}^2 \:+\: 
     \omega^2 \: {\varphi}^2 \: \right) 
   \; , $$ 
if $\varphi$ is the elongation, $m$ the mass, and $\omega$ the frequency of the oscillator.
Therefore the momentum is: 
   $$ \psi \;=\; 
      \frac{\partial\,u}{\partial\,\dot{\varphi}}
      \;=\; m \; \dot{\varphi}
      \; , $$ 
hence: 
   $$ u 
   \; =\; 
       \frac{m}{2} \: \omega^2 \: {\varphi}^2 
   \:+\: 
       \frac{1}{2\:m} \: {\psi}^2 
   \; . $$
From this it follows for the differential of the elementary domain according to (\ref{label_Planck15a_eq_17}): 
\vspace*{0mm}
\begin{align}
d\,G & \: = \; 
\bigintsss_{\;u}^{\,u+du}\!
d\varphi \;.\; d\psi
\;=\; 
\frac{2\:\pi}{\omega} \; du
\; ,
\nonumber 
\end{align}
for the energy limit $u_n$ according to (\ref{label_Planck15a_eq_4}): 
\vspace*{0mm}
\begin{align}
\bigintsss_{\;u=0}^{\,u=u_n}\!
d\,G & \: = \; \frac{2\:\pi}{\omega} \; u_n
\;=\; n \: h \; , 
\quad\quad 
u_n \;=\; \frac{n\:h\:\omega}{2\:\pi}
\; ,
\nonumber 
\end{align}
for the average energy in the elementary domain $n$ according to (\ref{label_Planck15a_eq_16}): 
\vspace*{0mm}
\begin{align}\tag{{\color{blue}24bis}}
 \overline{u}_n \;=\;
\frac{1}{h} \:  
\bigintsss_{\;u_n}^{\,u_{n+1}}\!
 u \:.\: d\,G & \: = \; 
\frac{h\:\omega}{2\:\pi} 
\left( n \:+\: \frac{1}{2} \,\right)
\; ,
\nonumber 
\label{label_Planck15a_eq_24bis}
\end{align}
and for the total energy in the stationary state according to (\ref{label_Planck15a_eq_19}): 
\vspace*{0mm}
\begin{align}
 U  & \: = \; 
 N \:.\: \frac{h\:\omega}{2\:\pi} 
\left( \frac{1}{\exp(\alpha) \:-\: 1 } 
       \:+\: \frac{1}{2} \,\right)
\; ,
\label{label_Planck15a_eq_24}
\end{align}
where 
$$ \alpha \;=\; \frac{h\:\omega}{2\:\pi\:k\:T} \; , $$ 
as {\it\color{blue}is} well-known.



\section{\underline{-- Planck (1915a) -- §\,6 {\it\color{blue}-- Rigid molecule around a fixed axis}} (p.414-415)}
\label{Section-17a-6}
\vspace*{-2mm}

For \dashuline{a rigid molecule that can be rotated about a fixed axis}, $f=1$, so again $G_n=h$, $p_n=1$, further: 
   $$ u \;=\; \frac{J}{2} \: \dot{\varphi}^2
        \;=\; \frac{J}{2} \: {\omega}^2
   \; , $$ 
where $\varphi$ is the angle position, $\omega$ the angular velocity, and $J$ the moment of inertia.
So: 
$$ \psi \;=\; \frac{d\,u}{d\,\omega} \;=\; J \: \omega \; ,$$ 
and therefore 
$$ u \;=\; \frac{1}{2} \: \frac{{\psi}^2}{J} \; . $$
From this it follows for the differential of the elementary domain according to (\ref{label_Planck15a_eq_17}):
\vspace*{0mm}
\begin{align}
d\,G & \: = \; 
\bigintsss_{\;u}^{\,u+du}\!
d\varphi \;.\; d\psi
\;=\; 2\:\pi \; 
   \sqrt{\frac{J}{2}} \:\; \frac{du}{\sqrt{u}}
\; ,
\nonumber 
\end{align}
for the energy limit $u_n$ according to (\ref{label_Planck15a_eq_4}): 
\vspace*{0mm}
\begin{align}
\bigintsss_{\;u=0}^{\,u=u_n}\!
d\,G & \: = \; 2\:\pi \; \sqrt{\:2\;J\;u_n}
\;=\; n \: h \; , 
\quad\quad 
u_n \;=\; \frac{n^2\:h^2}{8\:\pi^2\:J}
\; ,
\nonumber 
\end{align}
for the average energy in the elementary domain $n$ according to (\ref{label_Planck15a_eq_16}): 
\vspace*{0mm}
\begin{align}
 \overline{u}_n \;=\;
\frac{1}{h} \:  
\bigintsss_{\;u_n}^{\,u_{n+1}}\!
 u \:.\: d\,G & \: = \; 
\frac{h^2}{8\:\pi^2\:J} 
\left( n^2 \:+\: n \:+\: \frac{1}{3} \,\right)
\; ,
\nonumber 
\end{align}
and for the characteristic function $\Psi$ in the steady state according to (\ref{label_Planck15a_eq_20}): 
\vspace*{0mm}
\begin{align}
 \Psi  & \: = \; 
 N \:.\: k \:.\: 
\ln\!\left\{\;
\sum_{n=0}^{\infty} \;
\exp\left[\:
-\:\left( 
n^2 \:+\: n \:+\: \frac{1}{3} 
\right)
\: \sigma
\:\right]
\;\right\}
\; ,
\label{label_Planck15a_eq_25}
\end{align}
where 
\vspace*{-3mm}
\begin{align}
\sigma \;=\; \frac{h^2}{8\:\pi^2\:J\:k\:T} 
\; .
\label{label_Planck15a_eq_26}
\end{align}
This result is also 
well-known.\footnote{$\:$E. Holm, Ann. d. Phys. (4) {\bf 42}, 1311, 1913.}


\section{\underline{-- Planck (1915a) -- §\,7 {\it\color{blue}-- Rigid straight line around a fixed point}} (p.415-417)}
\label{Section-17a-7}
\vspace*{-2mm}

For \dashuline{a rigid straight line rotating freely around a fixed point}, $f=2$, so by (\ref{label_Planck15a_eq_5}): 
 $$ G_n \;=\; \left(\, 2\:n\:+\:1 \,\right) \; h^2 \; , $$ 
and by (\ref{label_Planck15a_eq_9}): 
 $$ p_n \;=\; 2\:n \:+\: 1 \; , $$ 
further: 
   $$ u \; =\; \frac{J}{2} \: 
     \left[\: \dot{\vartheta}^2 \:+\: 
     \sin^2(\vartheta) \; \dot{\varphi}^2 
     \:\right] 
   \; , $$ 
where $J $ is the main moment of inertia, and $\vartheta$ and $\varphi$ are the two polar coordinate angles of the straight line. 
The corresponding momentum coordinates are: 
$$ \eta \;=\; \frac{\partial\,u}{\partial\,\dot{\vartheta}} 
        \;=\; J \: \dot{\vartheta}  
   \quad\quad\mbox{and}\quad\quad  
   \psi \;=\; \frac{\partial\,u}{\partial\,\dot{\varphi}} 
        \;=\; J \: \sin^2(\vartheta) \; \dot{\varphi} 
   \; , $$ 
therefore:  
\vspace*{-2mm}
$$ u \; = \; \frac{1}{2\:J} \;.\; 
  \left( 
     \eta^2 \:+\: \frac{\psi^2}{\sin^2(\vartheta)} 
  \,\right) \; . $$
  
From this follows for the differential of the elementary region according to (\ref{label_Planck15a_eq_17}): 
\vspace*{0mm}
\begin{align}
d\,G & \: = \; 
\bigintsss\!\!\!
\bigintsss\!\!\!
\bigintsss\!\!\!
\bigintsss_{\;u}^{\,u+du}\!
d\vartheta \;.\; d\varphi \;.\; d\eta \;.\; d\psi
\;=\; 8\:\pi^2 \; J \:.\: du
\; ,
\nonumber 
\end{align}
for the energy limit $u_n$ according to (\ref{label_Planck15a_eq_4}): 
\vspace*{0mm}
\begin{align}
\bigintsss_{\;0}^{\,u_n}\! d\,G 
& \: = \; 8\:\pi^2\;J\:.\:u_n
\;=\; (\,n \: h \,)^2 \; , 
\quad\quad 
u_n \;=\; \frac{n^2\:h^2}{8\:\pi^2\:J}
\; ,
\nonumber 
\end{align}
for the mean energy in the elementary region $n$ according to (\ref{label_Planck15a_eq_16}): 
\vspace*{0mm}
\begin{align}
 \overline{u}_n \;=\;
\frac{1}{(2\:n\:+\:1)\:h^2} \:  
\bigintsss_{\;n}^{\,{n+1}}\!
 u \:.\: d\,G & \: = \; 
\frac{h^2}{8\:\pi^2\:J} 
\left( n^2 \:+\: n \:+\: \frac{1}{2} \,\right)
\; ,
\nonumber 
\end{align}
and for the characteristic function $\Psi$ in the stationary state according to (\ref{label_Planck15a_eq_20}): 
\vspace*{0mm}
\begin{align}
\boxed{\;
 \Psi \; = \; 
 N \:.\: k \:.\: 
\ln\!\left\{\;
\sum_{n=0}^{\infty} \;
\:(2\:n\:+\:1)\;.\:
\exp\left[\:
-\:\left( 
n^2 \:+\: n \:+\: \frac{1}{2} 
\right)
\: \sigma
\:\right]
\;\right\}
\;}
\; ,
\label{label_Planck15a_eq_27}
\end{align}
where $\sigma$ again has the meaning (\ref{label_Planck15a_eq_26}).

The expression (\ref{label_Planck15a_eq_27}), in conjunction with (\ref{label_Planck15a_eq_22}), gives the proportion $C_r$ of the heat capacity of a diatomic gas at constant volume that is due to the rotational energy.
This quantity {\it\color{blue}($\,C_r$)} differs in a characteristic way from that which Holm (l.c.) derived by simply doubling the expression (\ref{label_Planck15a_eq_25}) valid for a single degree of freedom.

While at high temperatures ($\sigma$ small) --for which the sums $\sum$ can be written as integrals and evaluated directly-- the heat capacity according to (\ref{label_Planck15a_eq_27}) actually becomes equal to the heat capacity according to the doubled expression (\ref{label_Planck15a_eq_25}), namely equal to $N\:k$ and corresponding to the two degrees of freedom, at sufficiently low temperatures ($\sigma$ large) the ratio of the former quantity to the latter becomes equal to $3/2$.

For large values of $\sigma$ it follows from (\ref{label_Planck15a_eq_22}) and according to the expression (\ref{label_Planck15a_eq_27}): 
$$ C_r \;=\; N \;.\; 12 \; k \; \sigma^2 
       \;.\; \exp(\, -\:2\;\sigma \,) 
\; , $$ 
with the {\bf\dashuline{zero-point energy}}: 
\vspace*{-3mm}
$$ N \;.\;\: \frac{h^2}{16\:\pi^2\:J} 
\; , $$ 
but according to the doubled expression (\ref{label_Planck15a_eq_25}) only: 
$$ C_r \;=\; N \;.\; 8 \; k \; \sigma^2 
       \;.\; \exp(\, -\:2\;\sigma \,) 
\; , $$ 
with a larger {\it\color{blue}(\,by the factor $16/12=4/3$)} {\bf\dashuline{\:zero-point energy}}: 
\vspace*{0mm}
$$ N \;.\;\: \frac{h^2}{12\:\pi^2\:J} 
\; . $$

According to his formula (25), Holm has the value of the constant $\sigma\:T=285$ for hydrogen from Eucken's 
measurements$\,$\footnote{$\:$Sitzungsber. d. peuss. Akad. d. Wiss. 1912, S.141}
and from this, according to (\ref{label_Planck15a_eq_26}), the moment of inertia $J=1.36.10^{-40}$ can be 
calculated.

When using formula (\ref{label_Planck15a_eq_27}), on the other hand, according to the measurements at low temperatures, the moment of inertia is slightly (but only very slightly, smaller) because the value of $C_r$ is very sensitive to small changes in $J$ when $\sigma$ is large.
The temperature range in which the two formulas diverge most, i.e. can be compared most closely, is unlikely to be very far below room temperature $(\sigma)$ for hydrogen.

Unfortunately, it seems that the $\sum$ in (\ref{label_Planck15a_eq_27}) cannot be reduced as easily as that in (\ref{label_Planck15a_eq_25}) to Jacobian theta functions, and thus to series that converge very quickly even with smaller $\sigma$.


\section{\underline{-- Planck (1915a) -- §\,8 {\it\color{blue}-- Impacts of Radiation. Conclusions}} (p.417-418)}
\label{Section-17a-8}

The theory developed above is limited to the determination of all properties of the stationary state, but it does not provide an answer to the question of which processes (and according to which laws) the stationary state is achieved, if an entirely arbitrary state initially prevails.

During the transition toward the stationary final state, very different types of processes can be effective, e.g. the mutual collisions of the molecules, or the effect of the electromagnetic field of thermal radiation (if you imagine the molecules to be electrically charged).
By assuming electric charges, the energy distribution in the stationary state cannot be modified in any way, as long as the molecules are all the same and the expression for the energy of a molecule remains the same.

If the laws of electrodynamic forces can now be assumed to be known, this results in a new 
way,\footnote{$\:$A. Einstein, Ann. d. Phys. (4) {\bf 17}, 132, 1905.} 
completely different from the previous one, to derive the conditions of the stationary state of the molecules
  by assuming the nature of the radiation field at a certain temperature as given by experience. In this derivation one does not need to use the concept of entropy and that of temperature at all.
The comparison of the result obtained in this way with that derived directly from the thermodynamic path described above then forms a very sensitive test of the admissibility of the theory developed.

In fact, in all the cases discussed here, the electrodynamic method gave exactly the same result as the thermodynamic one, and it is precisely this agreement that leads me to consider the hypotheses on which the above calculation is based to be correct.
However, the type of electrodynamic calculation is mathematically somewhat more complicated.
I have already done it for 
oscillators$\,$\footnote{$\:$Sitzungsber. d. preuss. Akad. d. Wiss. 1915, S.512.} 
and for dipoles with a fixed axis of 
rotation$\,$\footnote{$\:$Festschrift für Elster und Geitel, S.313, 1915.}.
For dipoles with free axes of rotation, I will share them on another occasion.

Conversely, if both theories (the thermodynamic and the electrodynamic) are assumed to be correct, the laws of blackbody radiation can be derived in this way, using an arbitrarily selected system of electrically charged molecules of any kind, by answering the question of how the radiation must be constituted so that its electrodynamic effects put the molecules in exactly the same state as that which results thermodynamically from the maximum entropy. This idea was also the basis of my first derivation of the law of radiation. 

I think I will soon deal 
{\color{blue}\citep[{\color{black}in a second communication}, by][]{Planck_Verh_dtsch_Phys_Ges_1915c}}
with the general case that the boundaries of the elementary areas of probability are not represented by the hypersurfaces of constant energy.

\vspace*{-2mm}
\begin{center}
---------------------------------------------------
\end{center}
\vspace*{-3mm}


\vspace*{-2mm}

\newpage
\begin{center}
\underline{\Large\bf Remarks on the entropy constant of diatomic gases}
\vspace*{2mm}

{\large\bf by Max 
 \citet[19 November][]{Planck_Verh_dtsch_Phys_Ges_1915b}}
\end{center}
\vspace*{-4mm}

\setcounter{section}{13} 
\setcounter{equation}{0}
\setcounter{figure}{0}  

\section{\underline{* Planck (1915b) {\it\color{blue}-- (a sole Section)}} (p.418-419)}
\label{Section-17b}
\vspace*{-2mm}

Mr. A. Tetrode, in a recent 
work,\footnote{$\:$H. Tetrode, Ber. d. Akad d. Wiss. v. Amasterdam v. 27. Februar 1915. 
{\it\color{blue} / 
In fact: ``\,Theoretical determination of the entropy constant of gases and liquids\,,'' KNAW, Proceedings, 17 III, 1914-1915, Amsterdam, 1915, pp.1167-1184
(\url{https://dwc.knaw.nl/DL/publications/PU00012754.pdf}) / 
Note also that Max Planck did not mention the previous papers of H. Tetrode (1912) and O. Sackur (1911, 1912, 1913) / P. Marquet.}} 
which was first brought to my attention by Mr. Einstein, developed expressions for the entropy of both monatomic and diatomic gases.
They differ from each other in that the entropy of a diatomic gas also contains something that comes from the rotational energy of the molecules.
So if you subtract the two expressions from each other, you get that part of the entropy of a diatomic gas that is determined by the rotations of the molecules, and the same is represented by the difference of the Tetrode's expressions (17) and 
(16):{\color{blue}\footnote{$\:$
{\it\color{blue}Note that, here: $N$ is the Avogadro's constant;  $k$ is the Planck-Boltzmann's constant; $h$ is the Planck's constant, and $J$ is the moment of inertia of the molecule / P. Marquet.}}} 
\begin{align}
\dbox{\ensuremath{\displaystyle\; 
S \; = \; 
k \; N \: 
\left\{\;
\ln(k\:T) \:+\: \ln(2\:\pi\:J) 
\:-\: 2\:\ln(h) \:+\: \ln(4\:\pi) 
\:+\: 1
\;\right\}
\;}}
\nonumber 
\:\; .
\end{align}

For the same quantity, a certain value also results from the theory I recently developed, with the help of an expression that I calculated as an example for a molecule with two degrees of  
freedom,\footnote{$\:$Max Planck, Verh. d. D. Phys. Ges. {\bf 17}, Sitzungster. vom. 5. November 1915. Formel 27), S.~416
{\it\color{blue}/ See Eq.~(27), p.416 in the previous paper by  \citet[][]{Planck_Verh_dtsch_Phys_Ges_1915a} / P. Marquet}.} 
namely: 
\begin{align}
\boxed{\;
\Psi \; = \; -\:\frac{F}{T} \; = \;
N \; k \: 
\ln\left\{\;
\sum_{n=0}^{\infty}
\: \left( 2\:n\:+\:1 \right)
\: . \:
\exp\left[\;
- \: \left( n^2\:+\:n\:+\:\frac{1}{2} \right)
  \: \sigma
  \;\right]
\;\right\}
\;}
\nonumber 
\: .
\end{align}
Here $F$ is the free 
energy$\,${\color{blue}\footnote{\label{label_footnote_Masieu}$\:${\it\color{blue}Note that the notation and the meaning of $\Psi$ correspond to the first of the two ``\,characteristic functions\,''
$\psi=S-U/T$ and $\psi'=S-(U+P\:V)/T$
introduced first by the French scientist Fran\c{c}ois Massieu in 1869, thus before that the (unnamed) functions 
$\psi=\epsilon-t\:\eta$ (for $F=U - T\:S$) and  
$\zeta=\epsilon+p\:v-t\:\eta$ (for $G=U +P\:V - T\:S$)
will be introduced by Gibbs (1875-78), and also before that the free-energy function 
$F=U - T \: S$ will be introduced (and named) by Helmholtz (1882-83). Max Planck has used and studied the Massieu's function  $\,\Psi=-\,F/T$ in all his German papers and book, with the interest that they are modified entropy functions and, as such, are more suitable for direct studies of maximum entropy states, for instance / P. Marquet.}}},
and 
\vspace*{0mm}
\begin{align}
\sigma & \: = \; 
\frac{h^2}{8\:\pi^2\:J\:k\:T}
{\color{magenta} 
\:\;=\; \frac{A}{T}
\;\;\;\mbox{where}\;\;\;
A \;=\; \frac{h^2}{8\:\pi^2\:J\:k}
}
\nonumber 
\; .
\end{align}

\dashuline{\,At high temperature} $(\sigma \ll 1)$, for which Tetrode's expression alone applies, the sum can be written as an 
integral,{\color{blue}$\,$\footnote{$\:${\it\color{blue}With the use of $-\,(n^2+n+1/2)=-\,(n+1/2)^2-1/4$,  $x=n+1/2$ and $dn=dx$ / P. Marquet}.}}
namely: 
\vspace*{0mm}
\begin{align}
 F & \:\: = \; 
 -\:T \; N \; k \: 
\ln\left\{\;
\bigintsss_{\;0}^{\,\infty}\!
\: 2\: \left( n\:+\:\frac{1}{2} \right)
\: . \:
\exp\left[\;
- \: \left( n \:+\:\frac{1}{2} \right)^2
  \: \sigma
  \;\right]
\: . \:
\exp\left( -\:\frac{\sigma}{4} \right)
\: . \:\; dn 
\;\right\}
\nonumber 
\; , \\ 
{\color{magenta} F}
& 
{\color{magenta}
\:\:\: = \; 
 -\:T \; N \; k \: 
\ln\left\{\;
\frac{-1}{\sigma} \; 
\bigintsss_{\;0}^{\,\infty}\!
\: (\: -\,2 \; \sigma \; x \:) \:
\exp\left[\;
- \: \sigma \: x^2 
  \;\right]
\: dx
\;\right\}
 \;-\; T \; N \; k \:
 \left[\;\frac{-\,\sigma}{4}\;\right]
}
\nonumber 
\; , \\ 
{\color{magenta} F}
& 
{\color{magenta}
\:\:\: = \; 
 -\:T \; N \; k \: 
 \left[\;
\ln\left(\;
\frac{1}{\sigma} 
\;\right)
\;\right]
 \;+\; T \; N \; k \:
 \left[\;\frac{\sigma}{4}\;\right]
}
\nonumber 
\; , \\ 
 F & \:\: = \; \;\;\;
 T \; N \; k \:
 \left[\:
  \frac{\sigma}{4} \:+\: \ln(\sigma)
 \:\right]
{\color{magenta} 
\:\;=\;  N \; k \:
 \left[\:
  \frac{A}{4} \:-\: 
  T \: \ln\left(\frac{T}{A}\right)
 \:\right]
}
\nonumber 
\; . 
\end{align}

If we now consider that{\color{blue}$\,$\footnote{$\:${\it\color{blue}With the use of $e=\exp(1)$ / P. Marquet}.}}
\vspace*{0mm}
\begin{align}
S & \: = \; 
-\:\frac{\partial\,F}{\partial\,T}
{\color{magenta} 
\:\;=\; 
 -\:N \; k \:
 \left[\:
  \ln\left(\sigma\right)
  \:-\: 1
 \:\right]
\;=\; 
 N \; k \:
 \left[\:
  \ln\left(\frac{1}{\sigma}\right)
  \:+\: 
  \ln\left({e}\right)
 \:\right]
\;=\; 
 N \; k \:
  \ln\left(\frac{e}{\sigma}\right)
}
\nonumber 
\; ,
\end{align}
substitution of the value of $\sigma$ results {\it\color{magenta}(at high temperatures)} in: 
\vspace*{0mm}
\begin{align}
& S \; = \; 
 N \; k \:
 \ln\left(\:
  \frac{8\:\pi^2\:J\:k\:T\:e}{h^2}
 \:\right)
\nonumber 
\; . \\
{\color{magenta} 
\mbox{or equivalently:}\;\;\;} & 
{\color{magenta}\dbox{\ensuremath{\displaystyle\; 
S \:\: = \; 
k \; N \: 
\left\{\;
\ln(k\:T) \:+\: \ln(2\:\pi\:J) 
\:-\: 2\:\ln(h) \:+\: \ln(4\:\pi) 
\:+\: 1
\;\right\}
} \:}}
\nonumber 
\: 
\end{align}
This is exactly the Tetrode's value {\it\color{magenta}(valid at high temperatures)} given above.

It goes without saying that this agreement makes each of the two independently developed theories more reliable, and thus provides further support for the applications of \dashuline{Nernst's heat 
theorem}{\color{blue}$\,$\footnote{$\:${\it\color{blue}Note that the ``\,Nernst's (1906) heat theorem\,'' was expressed in terms of 
``\,$\lim dA/dT = \lim dQ/dT = 0$\,'' assumed to be valid in the vicinity of $T=0$~K, whereas Planck rather considered the Third-law of thermodynamics he derived on his own as a generalisation of the Nernst's theorem: i.e. that the entropy (itself) is assumed to cancel out at $T=0$~K.
It must be emphasized here that the Tedrode-Planck formula is only valid at high temperature and cannot be used to calculate $S$ for $T$ close to $0$~K, where $S$ seems to tend towards minus infinity...
In fact, for very low temperatures, we must return to the formula for $\Psi$ with the sum for $n$ varying from $0$ to infinity, with $\sigma \propto 1/T \gg 1$ used to retain only the first term $(n=0)$, which leads to 
$F=-\,T\:\Psi = N \: k \: (\, T \: \sigma \,) \, / \, 2 
= N \: k \: A \, / \, 2= cste$, 
and thus to $S=-\,\partial F/\partial T=0$
/ P. Marquet}.}}.

\vspace*{-2mm}
\begin{center}
---------------------------------------------------
\end{center}
\vspace*{-3mm}


\newpage
\begin{center}
\underline{\Large\bf The quantum hypothesis for molecules with multiple degrees of freedom.}
\vspace*{0.1mm} \\ 
\underline{\Large\bf Second messages (final)}
\vspace*{2mm}

{\large\bf by Max 
 \citet[30 December][]{Planck_Verh_dtsch_Phys_Ges_1915c}}
\end{center}
\vspace*{-4mm}

\setcounter{section}{14} 
\setcounter{equation}{27}
\setcounter{figure}{0}  

\section{\underline{** Planck (1915c) -- Introduction} (p.438-439)}
\label{Section-17c-Introduction}
\vspace*{-2mm}

In the first communication {\it\color{blue}\citep[cf.][]{Planck_Verh_dtsch_Phys_Ges_1915a}}$\,$\footnote{$\:$Verh. d. D. Phys. ges. {\bf 17}, 407-418, 1915}  on the above topic, I limited myself to the treatment of the simplest case that in the $2\:f$-dimensional ``state space\,'' of a molecule formed by the general coordinates and the associated momenta, those surfaces which delimit the ``elementary regions of probability\,'' are all also surfaces of constant energy, i.e. that all states of the molecules corresponding to one and the same energy also lie in one and the same elementary region.
This condition is always fulfilled for a single degree of freedom.
For several degrees of freedom, however,  {\it\color{blue}(it is fulfilled)} only in special cases, such as the case of a rigid straight line 
that can rotate freely 
about a fixed point (§\,\ref{Section-17a-7}), since here the probability of a state can only depend on the magnitude of the rotational speed (i.e. on the energy), but not on the direction of the axis of rotation.

But if we take, for example, a 
point oscillating in a certain plane around a fixed equilibrium position (which also has two degrees of freedom), then, according to the quantum hypothesis and in contrast to the classical theory, the probability of any state depends not only on the energy, but also on the shape of the trajectory curve, and the family of surfaces of constant energy is no longer suitable for determining the elementary areas of probability.

This communication will now deal with the general case. Since it represents a direct continuation of the earlier one, the close connection is also expressed externally in that the meaning of the designations remains the same throughout, and the numbering of the paragraphs {\it\color{blue}(from §\,\ref{Section-17c-9} to §\,\ref{Section-17c-13})} and equations {\it\color{blue}(from \ref{label_Planck15c_eq_28} to \ref{label_Planck15c_eq_70})} is done in a continuous series.

\setcounter{section}{8} 
\section{\underline{** Planck (1915c) -- §\,9  {\it\color{blue}-- Characterization of States}} (p.439-440)}
\label{Section-17c-9}
\vspace*{-2mm}

The basic idea of the quantum hypothesis can be summarised, according to the view I have held for a long time, in the proposition that the thermodynamic probability of each state can be expressed by a certain integer, and this necessarily means that in the state space of a molecule formed by the general coordinates and momenta, the elementary regions on which the calculation of the probability must be based have a very specific position, size and shape, or in other words that the state space has a very specific structure. 
The task that then remains to be solved consists solely in defining those elemental areas.
If these are known, the entropy and temperature for a large number of similar molecules in the stationary state can be clearly calculated using the methods described in §\,\ref{Section-17a-2} and §\,\ref{Section-17a-3}.

Now, it was already mentioned in the introduction that a single set of surfaces 
\vspace*{0mm}
\begin{align}
 g \; = \; 0
 \;,\quad\quad
 g \; = \; g_1
 \;,\quad\quad
 g \; = \; g_2
 \;,\quad .\:.\:. \quad
 g \; = \; g_n
 \;,\quad .\:.\:. 
\label{label_Planck15c_eq_28}
\end{align}
is generally not sufficient to determine the boundaries of the elementary regions in the state space of a molecule.
Rather, the space between two adjacent surfaces $g=g_n$ and $g=g_{n+1}$ can also include many elementary areas.
For this reason, we introduce further surface sets for the general case: 
\vspace*{0mm}
\begin{align}
 g' \; = \; 0
 \;,\quad\quad\:
 g' \; = \; g'_1
 \;,\quad\quad\:
 g' \; = \; g'_2
 \;,\quad\; 
 .\:.\:. \quad\,
 g' \; = \; g'_q
 \;,\quad .\:.\:. 
\nonumber \\ 
 g'' \; = \; 0
 \;,\quad\quad
 g'' \; = \; g''_1
 \;,\quad\quad
 g'' \; = \; g''_2
 \;,\quad 
 .\:.\:. \quad
 g'' \; = \; g''_r
 \;,\quad .\:.\:. 
\label{label_Planck15c_eq_29} \\ 
 .\;\;\;.\;\;\;.\;\;\;.\;\;\;.\;\;\;.\;\;\;.\;\;\;.\;\;\;.\;\;\;.\;\;\;.\;\;\;.\;\;\;.\;\;\;.\;\;\;.\;\;\;.\;\;\;.\;\;\;.\;\;\;.\;\;\;.\;\;\;.\;\;\;.\;\;\;.\;\;\;.\;\;\;.
\nonumber 
\end{align}
The quantities $g,\:g',\:g'', \:.\,.\,.$ are certain functions of the general coordinates and momenta, the quantities $g_n,\:g'_q,\:g''_r, \:.\,.\,.$ are certain constants that depend only on the nature of the molecules.
All these surfaces divide the entire $2\,f$-dimensional state space of a molecule into an infinite number of \:``\/cells\,'' corresponding to the number of pairs of surfaces, and each such cell bounded by two neighbouring pairs of surfaces from each group represents an elementary region of probability.
Therefore, to characterise a certain elementary area $(n\:q\:r\:.\:.\:.)$, let us consider the one bounded by the surfaces $g=g_n$, $g=g_{n+1}$, $g'=g'_q$, $g'=g'_{q+1}$, $g''=g''_r$, $g''=g''_{r+1}$, \:.\:.\:. 
Then every point of the state space (i.e. every value system of coordinates and impulsions) lies in a certain elementary area.


\section{\underline{** Planck (1915c) -- §\,10 {\it\color{blue}-- Computation of $\,g, \:g', \:g'', \:.\:.\:.$} } (p.440-442)}
\label{Section-17c-10}
\vspace*{-2mm}

It is now primarily a matter of determining the functions $g, \:g', \:g'', \:.\:.\:.$ 
For this purpose: \\ \dashuline{a first condition} is provided by the theorem that all these quantities $g$ (as well as the energy $u$) retain constant values if the state point of the molecules under consideration in the state space passes through the curve prescribed by the laws of dynamics if the molecules are left to their own devices.

In other words, the quantities $g, \:g', \:g'', \:.\:.\:.$ are made up of the integration constants $u, \:v, \:w, \:.\:.\:.$ of the differential equations of the state curve: 
$$
d\,\varphi_1 : d\,\varphi_2 
\,:\, \;.\:.\:.\; \,:\, 
d\,\psi_1 : d\,\psi_2 
\,:\, \;.\:.\:.\;
      \;\;=\;\;\;
\frac{\partial\,u}{\partial\,\psi_1} : 
\frac{\partial\,u}{\partial\,\psi_2} 
\,:\, \;.\:.\:.\; \,:\, 
-\:\frac{\partial\,u}{\partial\,\varphi_1} : 
-\:\frac{\partial\,u}{\partial\,\varphi_2} 
\,:\, \;.\:.\:.\;
$$
Because in a process that takes place according to purely dynamic laws, the probability remains constant under all circumstances, so the state point cannot possibly pass through an interface between two elementary areas of probability in the course of its movement.

\dashuline{A second condition} is the consideration that the position of the interfaces $g=const.$, because of their physical meaning, cannot depend on the choice of coordinate system, in particular on the direction of the coordinate axes.

\dashuline{Thirdly}, it must be required that the elementary areas of probability completely fill the entire available area of the $u, \:v, \:w, \:.\:.\:.$ (without gaps), i.e. that a finite limit of this area (that is fixed from the outset) is in any case one of the boundary areas of the elementary areas.

\dashuline{As a fourth condition} for finding the functions $g, \:g', \:g'', \:.\:.\:.$ and at the same time for calculating the constants $g_n, \:g'_q, \:g''_r, \:.\:.\:. $, in generalisation of (\ref{label_Planck15a_eq_4}) the following equation 
\vspace*{0mm}
\begin{align}
G \;=\;
\bigintsss_{\;0}^{\,g_n}\!\!\!
\bigintsss_{\;0}^{\,g'_q}\!\!\!
\bigintsss_{\;0}^{\,g''_r}\!\!\!
\bigintsss_{\;.\,.\,.}^{\,.\,.\,.}
d\,G \;=\; 
\left( n \: h \right)^{i} \;.\: 
\left( q \: h \right)^{i'} \;.\: 
\left( r \: h \right)^{i''} \;.\:.\:. 
\; 
\label{label_Planck15c_eq_30}
\end{align}
is used for the size $G$ of the state space enclosed by the surfaces 
\vspace*{0mm}
\begin{align}
 g \; = \; 0
 \;,\quad
 g \; = \; g_n
 \;,\quad\quad
 g' \; = \; 0
 \;,\quad
 g' \; = \; g'_q
 \;,\quad\quad
 g'' \; = \; 0
 \;,\quad
 g'' \; = \; g''_r
 \;,\quad .\:.\:. 
 \;\;\; ,
\nonumber 
\end{align}
where 
\vspace*{-2mm}
\begin{align}
d\,G \;=\;
\bigintsss  \!\!\!
\bigintsss  \!\!\!
\:.\,.\,.\: \!\!\!
\bigintsss_{\;\;\;\;\;\;g,\;\;\;\;\;\;\;\;\;g',\;\;\;\;\;\:.\:.\:.}^{\,g+dg,\:g'+dg',\;\:.\:.\:.}\;\;
d\,\varphi_1 \;.\;
d\,\varphi_2 \;\;.\;.\;.\;\;
d\,\psi_1 \;.\;
d\,\psi_2 \;\;.\;.\;.\; 
\label{label_Planck15c_eq_31}
\; ,
\end{align}
and for the integers 
\vspace*{-2mm}
\begin{align}
i \;+\; i' \;+\; i'' \;+\; .\:.\:. \;=\; f
\label{label_Planck15c_eq_32}
\; .
\end{align}
In expression (\ref{label_Planck15c_eq_31}), which only depends on the quantities $\:g, \:g', \:g'', \:.\:.\:.\:$ and their differentials, the integration is
to be thought of as being  
carried out over all those points of the state space 
which lie within the infinitely thin layer of space designated by the limits of the integrals.
They of course all belong to the same elementary domain, but can otherwise 
  still form a manifold that is many times infinite,
since the number of quantities $\:g, \:g', \:g'', \:.\:.\:.\:$ 
can be 
considerably
smaller than that of the coordinates and impulsions.

The relationship (\ref{label_Planck15c_eq_32}) results directly from (\ref{label_Planck15c_eq_30}) by considering the dimensions of $G$ and $h$.
  Furthermore, (\ref{label_Planck15c_eq_30}) gives rise to the curious theorem that the $f$ degrees of freedom are distributed in a very specific way over the sets of surfaces $g$, in such a way that   to  each of the functions $\:g, \:g', \:g'', \:.\:.\:.\:$ is assigned a certain number of degrees of freedom.
  
As a result, the degrees of freedom break down into specific groups, and two degrees of freedom belonging to the same group can be described as ``\,coherent\,'' whereas two degrees of freedom belonging to different groups can be described as ``\,incoherent.\,''
For example, the two degrees of freedom of a rigid straight line rotating around a fixed point are coherent, because there is only a single set of surfaces $q$. In contrast, the two degrees of freedom of a plane oscillator are incoherent, because one belongs to the $g=const.$ set and the other to the $g'=const.$ set.

Since the ordinal numbers $\:n,\:q,\:r,\:.\:.\:.\:$ are independent of each other, in order to fulfill equation (\ref{label_Planck15c_eq_30}) it is necessary that it is possible through a suitable choice of Functions $\:g,\:g',\:g'',\:.\:.\:.\:$ to bring the differential $d\,G$ to the form: 
\vspace*{-3mm}
\begin{align}
d\,G \;=\; dg\;.\;dg'\;.\;dg'' \;.\:.\:.
\label{label_Planck15c_eq_33}
\end{align}
or after integration into the considered state space: 
\vspace*{0mm}
\begin{align}
G \;=\; g_n\;.\;g'_q\;.\;g''_r \;.\:.\:.
\label{label_Planck15c_eq_34}
\end{align}
Then equation (\ref{label_Planck15c_eq_30}) is satisfied if: 
\vspace*{-2mm}
\begin{align}
g_n \;=\; \left( n \: h \right)^{i} \;, \;\;\; 
g'_q \;=\; \left( q \: h \right)^{i'} \;, \;\;\;  
g''_r \;=\; \left( r \: h \right)^{i''} \;.
\label{label_Planck15c_eq_35}
\end{align}
However, this procedure for determining the functions $\:g,\:g',\:g'',\:.\:.\:.\:$ is not yet clear, because you could e.g. double the function $g$  in (\ref{label_Planck15c_eq_33}), and then halve the function $g'$. Then (\ref{label_Planck15c_eq_34}) remains unchanged, and one would get completely different interfaces from (\ref{label_Planck15c_eq_35}).

A method for separating the functions $\:g,\:g',\:g'',\:.\:.\:.\:$ is provided by \\ \dashuline{the following fifth condition}: 

If you imagine the freedom of movement of the molecules to be limited in such a way that $i'=0$, $i''=0$, \:.\:.\:.\: and only $i$ retains its value, 
then only a single set of surfaces $g=const$ remains to delimit the elementary areas, and we have the case discussed in the first communication, where simply $d\,G=dg$. 
However, the expression for $g$ calculated from this also retains its meaning for the general case, since the function $g$ is independent of the functions $\:g',\:g'',\:.\:.\:.\:$ 

Proceeding further, one then sets $i''$ and the following degrees of freedom equal to zero, and deals with the case of molecules equipped with $i+i'$ degrees of freedom.
Since $g$ is already known, $g'$ follows clearly from (\ref{label_Planck15c_eq_33}), and one continues in this way step by step until the general case is solved.
Each variation in the order of $g$ results in a means for testing the consistency of the theory. For the application of the procedure described, see the calculation carried out in the next §\,\ref{Section-17c-12}.



\section{\underline{** Planck (1915c) -- §\,11  {\it\color{blue}-- Computations of $\,G, \, U, \,S, \,\Psi$} } (p.442-444)}
\label{Section-17c-11}
\vspace*{-2mm}

With the determination of the functions $g$ the essential part of the task is done. The size of the elementary area $(n\:q\:r\:.\,.\,.\:)$ of the probability results, according to the definition in §\,\ref{Section-17c-9} and with the help of (\ref{label_Planck15c_eq_30}) {\it\color{blue}and $f$ given by  (\ref{label_Planck15c_eq_32})}, as follows:
\vspace*{0mm}
\begin{align}
G_{n\,q\,r\,.\,.\,.} \;=\;
\left[\;
  \left( n \,+\, 1 \right)^{i} 
  \:-\:
  \left( n \right)^{i} 
\;\right]\;.\: 
\left[\;
  \left( q \,+\, 1 \right)^{i'}
  \:-\:
  \left( q \right)^{i'}
\;\right]\;.\: 
\left[\;
  \left( r  \,+\, 1\right)^{i''}
  \:-\:
  \left( r \right)^{i''}
\;\right]
\:\; \:.\:.\:. \:\; h^f
\; .
\label{label_Planck15c_eq_36}
\end{align}
Since the elementary domain $(000\:...)$ has the size $h^f$, then
\vspace*{0mm}
\begin{align}
G_{n\,q\,r\,.\,.\,.} \;=\;
 p\;.\; G_{0\,0\,0\,.\,.\,.} 
\label{label_Planck15c_eq_37}
\end{align}
where $p$ is an integer, namely: 
\vspace*{0mm}
\begin{align}
p \;=\;
\left[\;
  \left( n \,+\, 1 \right)^{i} 
  \:-\:
  \left( n \right)^{i} 
\;\right]\;.\: 
\left[\;
  \left( q \,+\, 1 \right)^{i'}
  \:-\:
  \left( q \right)^{i'}
\;\right]
\:\; \:.\:.\:.
\label{label_Planck15c_eq_38}
\end{align}

According to these determinations, 
the calculation of the probability and entropy of any given state (of a large number $N$ of identical, independently vibrating molecules), as well as the search for the conditions of the stationary state and the calculation of the energy, entropy and free energy (as a function of temperature), can be carried out exactly according to the method discussed in §\,\ref{Section-17a-2} and §\,\ref{Section-17a-2}, so that we can limit ourselves here to stating/citing the main results.

{\it\color{blue}By generalizing  
(\ref{label_Planck15a_eq_12}) to 
(\ref{label_Planck15a_eq_14}),}
the entropy of any given state of the system is: 
\begin{align}
S & \: = \; 
-\:k \:.\: 
\sum_{n=0}^{\infty} \: 
\sum_{q=0}^{\infty} \: 
\:.\:.\:. \;\;
w_{\,n\,q\,.\,.\,.} 
\; \ln\left(\frac{w_{\,n\,q\,.\,.\,.}}{p} \right)
\label{label_Planck15c_eq_39}
\;\: ,
\end{align}
where: 
\vspace*{-3mm}
\begin{align}
w_{\,n\,q\,.\,.\,.} & \: = \; \frac{N_{n\,q\,.\,.\,.}}{N}
\label{label_Planck15c_eq_40}
\; , \: 
\end{align}
and if $N_{n\,q\,.\,.\,.}$ means the number of molecules that fall into the elementary region $\,(n\:q\: .\,.\,.)\,$ in the given state, i.e.: 
\vspace*{-3mm}
\begin{align}
\sum_{n=0}^{\infty} \: 
\sum_{q=0}^{\infty} \: 
\:.\:.\:. \;\;
w_{\,n\,q\,.\,.\,.} 
& \: = \; 1
\label{label_Planck15c_eq_41}
\;\: .
\end{align}
  Since the summations over the ordinal numbers $\,n,\,q,\,.\,.\,.\:$ are to be carried out in the same way everywhere, for the sake of simplicity they are generally replaced by a single $\sum$ sign, and the ordinal numbers $\,n,\,q,\,.\,.\,.\:$ are also omitted, as was the case with the number $p$ in (\ref{label_Planck15c_eq_38}).

In the stationary state of the system, the total energy
{\it\color{blue}(up to any constant reference term 
$U_{ref}=N\:\overline{u}_{ref}$)} is
\vspace*{-3mm}
\begin{align}
U \;=\; N \; \overline{u} \;=\;
\sum \:\; N_{\,n\,q\,.\,.\,.} \;\;
\overline{u}_{\,n\,q\,.\,.\,.} 
\label{label_Planck15c_eq_42}
\;\: ,
\end{align}
whereby the average energy in the elementary region $\,(\,n,\,q,\,.\,.\,.)\:$ according to (\ref{label_Planck15c_eq_31}) and (\ref{label_Planck15c_eq_37}) is 
\vspace*{0mm}
\begin{align}
\overline{u}_{\,n\,q\,.\,.\,.} 
  \;=\;
\bigintsss_{\;\;\;\;\;g_n,\;\;\;\;\;\;g'_q,\;\;\;\:.\:.\:.}^{\,g_{n+1},\:g'_{q+1},\;\:.\:.\:.}\;\;
u \;.\; d\,G \::\: ( p \: h^f ) \;\:
\label{label_Planck15c_eq_43}
\; .
\end{align}
From these results
{\it\color{blue}--and from (\ref{label_Planck15a_eq_18})--}
the distribution number (in the elementary region mentioned) is: 
\vspace*{0mm}
\begin{align}
w_{\,n\,q\,.\,.\,.}  & \: = \; 
\frac{\displaystyle 
p \:.\: \exp\left( 
-\,\frac{\overline{u}_{\,n\,q\,.\,.\,.}}{k\:T} \right)
}
{\displaystyle  
\sum \;
p \:.\: \exp\left( 
-\,\frac{\overline{u}_{\,n\,q\,.\,.\,.}}{k\:T} \right)
}
\label{label_Planck15c_eq_44}
\; ,
\end{align}
{\it\color{blue}--from (\ref{label_Planck15a_eq_19})--}
the energy as a function of temperature is: 
\vspace*{-2mm}
\begin{align}
U \:=\; N \;.\;\: \overline{u}
 & \: = \; 
 N \:.\:
\frac{\displaystyle 
\sum \;
p \; \overline{u}_{\,n\,q\,.\,.\,.} \; \exp\left( 
-\,\frac{\overline{u}_{\,n\,q\,.\,.\,.}}{k\:T} \right)
}
{\displaystyle  
\sum \;
p \; \exp\left( 
-\,\frac{\overline{u}_{\,n\,q\,.\,.\,.}}{k\:T} \right)
}
\label{label_Planck15c_eq_45}
\; ,
\end{align}
{\it\color{blue}--from (\ref{label_Planck15a_eq_19bis})--} the entropy: 
\vspace*{0mm}
\begin{align}
S \:=\; N \; \overline{s} 
  \:=\; N \:.\: \left\{\;
  \frac{\overline{u}}{T}
  \;+\;
  k \: \ln\left[\;
\sum \;
p \; \exp\left( 
-\,\frac{\overline{u}_{\,n\,q\,.\,.\,.}}{k\:T} \right)
\;\right]
\;\right\}
\label{label_Planck15c_eq_46}
\; ,
\end{align}
and finally 
{\it\color{blue}--from (\ref{label_Planck15a_eq_20})--} 
the free energy $F$ and the {\it\color{blue}(Massieu's)} characteristic function $\Psi$
{\it\color{blue}(see the footnote$\,{}{\mbox{\ref{label_footnote_Masieu}}}$)}: 
\vspace*{-3mm}
\begin{align}
\Psi \:=\; -\:\frac{F}{T}
\:=\; N \; k \; 
\ln\!\left[\;
\sum \;
p \; \exp\left( 
-\,\frac{\overline{u}_{\,n\,q\,.\,.\,.}}{k\:T} \right)
\;\right]
{\color{blue}\;\:=\; S \:-\: \frac{U}{T}\:
 \;=\; N \; k \; \ln\!\left(\: Z \:\right)
}
\label{label_Planck15c_eq_47}
\; ,
\end{align}
from which the heat capacity $C$ can also be calculated directly according to (\ref{label_Planck15a_eq_22}).

The form of these expressions is very similar to that derived in the previous communication {\it\color{blue}\citep{Planck_Verh_dtsch_Phys_Ges_1915a}} for purely coherent degrees of freedom (see §\,\ref{Section-17a-3}).
The only difference is that the summation has to be extended over several types of ordinal numbers. 

In order for the sum consisting of multiple infinite terms to have a finite value, it is necessary that $\overline{u}_{\,n\,q\,.\,.\,.}$ always grows to infinity as the ordinal numbers {\it\color{blue}($p$)} grow {\it\color{blue}(so that the exponentials of $-\,(\overline{u}_{\,n\,q\,.\,.\,.})/(k\:T)$ decrease to $0$ more  rapidly than $p$ increase)}, and this means that the expression of $\overline{u}_{\,n\,q\,.\,.\,.}$ contains all ordinal numbers.
Otherwise one of the ordinal numbers could grow to infinity, while the exponential value in the final equations would remain constant. 

The surfaces of constant energy therefore do not belong to the boundary surfaces $\:g,\:g',\:g'',\:.\:.\:.\:$ of the elementary regions, since each of these only depends on a single atomic number.

If the ``\,molecules\,'' have very many degrees of freedom $(f \gg 1)$, then each state of the molecules represents a point in Gibbs phase space, and the system of $N$ molecules under consideration plays the role in the quantum hypothesis that is assigned to the ``\,canonical set\,'' in classical theory. 
For high temperatures, where the summations can be replaced by integrations, the formulae in fact merge directly into those of the classical theory. 
For sufficiently low temperatures, the whole energy $U$ is reduced to that of the elementary region 
$(\,0\,0\,.\,.\,.)$, namely: 
$\,N\:\overline{u}_{\,0\,0\,.\,.\,.}$.



\section{\underline{** Planck (1915c) -- §\,12  {\it\color{blue}-- A mass point in a center of force} } (p.445-450)}
\label{Section-17c-12}
\vspace*{-2mm}

  We now want to apply the developed formulas to a case that is as important as it is instructive, namely that of a point with a  mass $m$ oscillating periodically in space around a fixed, attractive center of force. 
For this case, the number of degrees of freedom is $f=3$, since the state of the oscillator is determined by the three types of polar coordinates: 
\vspace*{1mm} \\ \hspace*{25mm}
$r\;(>0)$, 
\;\;\;\;$\vartheta$ (between $0$ and $\pi$) 
\;\;\;and \;$\varphi $ (between $0$ and $2\:\pi$) 
\vspace*{1mm} \\ \hspace*{0mm}
and the corresponding impulses: 
\vspace*{-3mm}
\begin{align}
\rho \;=\; m \; \dot{r} \; , 
\quad\quad
\eta \;=\; m \; r^2 \; \dot{\vartheta} \; , 
\quad\quad
\psi \;=\; m \; r^2 \: \sin^2(\vartheta) \;\: \dot{\varphi}
\label{label_Planck15c_eq_48}
\; .
\end{align}

The five conditions set out in §\,\ref{Section-17c-10} are used to find the functions $\:g,\:g',\:.\:.\:.\:$. 
Since according to the first condition the quantities $\:g,\:g',\:.\:.\:.\:$ are derived from the integration constants $\:u,\:v,\:.\:.\: .\:$ of the dynamic differential equations of the state curve, we first consider the state curve, i.e. the laws of free movement of the oscillator. 
This takes place in a plane and is determined by the position of this plane and by the constants of the energy and the rotational moment. 
Since, according to the second condition, the quantities $g$ do not depend on the directions of the coordinate axes, they do not depend on the position of the orbital plane, but only on the values of the energy $(u)$ and the rotational moment $(v > 0)$.
These two quantities 
\vspace*{-3mm}
\begin{align}
u \;=\; \frac{1}{2\:m} \:
\left( \: \rho^2 
 \;+\; \frac{\eta^2}{r^2} 
 \;+\; \frac{\psi^2}{r^2\;\sin^2(\vartheta)} 
\; \right)
\;+\; f(r)
\label{label_Planck15c_eq_49}
\; 
\end{align}
[where $f(r)$ is the potential energy] and 
\vspace*{0mm}
\begin{align}
v^2 \;=\; \eta^2
    \;+\; \frac{\psi^2}{\sin^2(\vartheta)} 
\label{label_Planck15c_eq_50}
\; 
\end{align}
thus correspond to two sets of interfaces $g=g_n$ and $g'=g'_q$, and of the three degrees of freedom, two are coherent with each other according to (\ref{label_Planck15c_eq_32}). 

We reserve the fulfillment of the third condition (concerning the boundaries of the area of $u$ and $v$) for later, as we want to leave the question of the law of attraction by which those boundaries are determined still open.

First, to satisfy the fourth condition, we express the differential $d\,G$ of the state space  with the following six new state variables: $u, \:v, \:\vartheta', \:\varphi', \:r, \:\chi$ (rather than through the six state variables $r, \:\vartheta, \:\varphi, \:\rho, \:\eta, \:\psi$).
  Here $\vartheta'$ (between $0$ and $\pi$) and $\varphi'$ (between $0$ and $2\:\pi$) are the polar coordinate angles of the positive normals of the orbital plane (which also determines the sense of the orbit), and $\chi$ (between $0$ and $2\:\pi$) is the angle of the radius vector $r$ with that fixed radius vector which is formed by the projection of the positive $z$-axis onto the orbital plane.
Then the relationships between the old and the new state variables are: 
\vspace*{-3mm}
\begin{align}
r & \:=\; r
\nonumber \; , \\ 
\cos(\vartheta) & \:=\; \sin(\vartheta')\;\:\cos(\chi)
\label{label_Planck15c_eq_51} \; , \\ 
\cos(\varphi \:-\: \varphi') & \:=\; 
-\:\frac{\cos(\vartheta')\;\cos(\chi)}
        {\sin(\vartheta)} 
\; , \quad
\tan(\varphi \:-\: \varphi') \;=\; 
\frac{\tan(\chi)}
     {\cos(\vartheta')} 
\label{label_Planck15c_eq_52} \; , \\ 
\rho^2 & \:=\; 2\;m\;u \;-\; \frac{v^2}{r^2}
         \;-\; 2\;m\;f(r)
\label{label_Planck15c_eq_53} \; , \\ 
\eta & \:=\; 
\frac{v\;\sin(\vartheta')\;\sin(\chi)}
     {\sin(\vartheta)} 
\label{label_Planck15c_eq_54} \; , \\ 
\psi & \:=\; v\;\cos(\vartheta')
\label{label_Planck15c_eq_55} \; , 
\end{align}
and the differential (\ref{label_Planck15c_eq_31}) of the state space becomes 
\vspace*{0mm}
\begin{align}
d\,G \;=\;
\bigintsss  \!\!\!
\bigintsss  \!\!\!
\bigintsss  \!\!\!
\bigintsss  \!\!\!
\bigintsss  \!\!\!
\bigintsss_{\;\;\;\;\;\;g,\;\;\;\;\;\;\;\;\;g'}^{\,g+dg,\:g'+dg'}\;\;
D \:\:.\:\: 
d\,u \:\:.\:\:
d\,v \:\:.\:\:
d\,\vartheta' \:\:.\:\:
d\,\varphi' \:\:.\:\: 
d\,r \:\:.\:\:
d\,\chi 
\label{label_Planck15c_eq_56}
\;\: ,
\end{align}
where the functional determinant is 
\vspace*{0mm}
\begin{align}
D \;=\; \pm \;\; 
\left|\; 
\frac{\partial\:r}{\partial\:u} 
\: , \:.\:.\:.\; , \;
\frac{\partial\:\psi}{\partial\:\chi} 
\;\right|
\;=\; \frac{m}{\rho} \:\; v \; \sin(\vartheta')
\label{label_Planck15c_eq_57}
\;\: ,
\end{align}
by which we understand $\rho$ to be the positive root of (\ref{label_Planck15c_eq_53}).

The integration of (\ref{label_Planck15c_eq_56}) is to be extended 
over $\vartheta'$ from $0$ to $\pi$, 
over $\varphi'$ from $0$ to $2\:\pi$, 
over $r$ from $r_{min}$ to $r_{max}$ (for certain $u$ and $v$), 
over $\chi$ from $0$ to $2\: \pi$, 
and in addition the whole expression must be multiplied by $2$, because due to the double value of $\rho$ in (\ref{label_Planck15c_eq_53}) the new state variables do not determine the state of the oscillator unambiguously, but ambiguously.


This results in: 
\vspace*{0mm}
\begin{align}
d\,G \;=\;
16\;\pi^2\;\: v \:\:.\:\:
           d\,u \:\:.\:\:
           d\,v \:\:.\:\:
\bigintsss_{\;r_{min}}^{\,r_{max}}\;\;
\frac{m\;d\,r}{\rho}
 \nonumber 
\;\: .
\end{align}
Then, the integral over $r$ according to (\ref{label_Planck15c_eq_53}) means nothing other than the time during which the oscillating point passes from the smallest to the greatest distance from the center of force.
If we denote it by $\tau$, we get: 
\vspace*{-3mm}
\begin{align}
d\,G \;=\; 8\;\pi^2 \;\:\tau \:\:.\:\:
           d\,u \:\:.\:\: d\,v^2
\label{label_Planck15c_eq_58}
\;\: .
\end{align}
We want to adapt the further treatment of the task to the case that the oscillator oscillates according to 
Hook's law{\color{blue}$\,$\footnote{$\:${\it\color{blue}The Hook's law (Wikipedia) states that the force $(F)$ needed to extend or compress a spring by some distance $(x)$ scales linearly with respect to that distance: that is, $F \, = \, k \: x$, where $k$ is a constant factor characteristic of the spring, and $x$ is small compared to the total possible deformation of the spring / P. Marquet}.}}.
Then the potential energy is: 
\vspace*{0mm}
\begin{align}
f(r) \;=\; \frac{m}{2}\;\:\omega^2\;\:r^2
\label{label_Planck15c_eq_59}
\;\: ,
\end{align}
where $\omega$ denotes the constant oscillation frequency. 
The time $\tau$ here forms the fourth part of the entire oscillation time, i.e. 
$$ \tau \;=\; \frac{1}{4} \;.\; \frac{2\:\pi}{\omega} 
 \;=\; \frac{\pi}{2\;\omega} \; ,$$ 
hence: 
\vspace*{-3mm}
\begin{align}
d\,G \;=\; \frac{4\;\pi^3}{\omega}  \:\:.\:\:
           d\,u \:\:.\:\: d\,v^2
\label{label_Planck15c_eq_60}
\;\: .
\end{align}
Although $d\,G$ 
    is given by 
(\ref{label_Planck15c_eq_33})
    {\it\color{blue} 
    (and thus $d\,G \;=\; dg\;.\;dg'$)}, 
the quantities $u$ and $v^2$ must not be taken to be proportional to $g$ and $g'$, 
because apart from the fact that, according to the remark about (\ref{label_Planck15c_eq_47}){\color{blue}$\,$\footnote{$\:${\it\color{blue}Namely that: ``\,The surfaces of constant energy therefore do not belong to the boundary surfaces $\:g,\:g',\:g'',\:.\:.\:.\:$ of the elementary regions, since each of these only depends on a single atomic number\,'' / P. Marquet}.}}  
that the energy $u$ 
   has nothing to do with 
the functions $g$ and $g'$, 
the fulfilment of \dashuline{the third condition} is still missing 
(namely that the finite boundaries of the entire domain of $u$ and $v$ necessarily belong to the areas $g=const$ and $g'=const$).

Now the area of $u$ and $v$ (in the finite) is limited firstly by the area $v=0$, which corresponds to the rectilinear oscillations (the equation $u=0$ does not give an area, but a point) and secondly by the area which expresses the boundary condition for the fact that the roots of the equation $\rho=0$ in $r$ are real. 
  This boundary condition of (\ref{label_Planck15c_eq_59}) is obtained from (\ref{label_Planck15c_eq_53}): 
$u-\omega\:v=0$. 
If we set this always positive quantity 
$$ u \;-\; \omega\:v \;=\; u' \; , $$ 
the area $u'=0$ denotes a second limit of the area of $u$ and $v$, i.e. the circular oscillations.

By introducing $u'$ and $v^2$ instead of $u$ and $v^2$ we get from (\ref{label_Planck15c_eq_60}) and (\ref{label_Planck15c_eq_33}): 
\vspace*{0mm}
\begin{align}
d\,G \;=\; dg\;.\;dg' 
     \;=\; \frac{4\;\pi^3}{\omega}  \:\:.\:\:
           d\,u' \:\:.\:\: d\,v^2
\nonumber 
\;\: ,
\end{align}
where now $u'$ and $v^2$ can be changed completely independently of each other between $0$ and $\infty$, and fill the entire state space with all their values without any gaps.

We can therefore set $g$ proportionally to $u'$, and $g'$ proportionally to $v^2$, and then get 
(\ref{label_Planck15c_eq_30}) and 
(\ref{label_Planck15c_eq_35}) with $f=3$: 
\vspace*{-3mm}
\begin{align}
G \;=\; \frac{8\;\pi^3}{\omega}  \:\:.\:\:
           u'_n \:\:.\:\: v^2_q
  \;=\; g_n\;.\;g'_q 
  \;=\; n\;h\;.\;(q\:h)^2 
\nonumber 
\;\: .
\end{align}
The two coherent degrees of freedom are assigned to the surfaces $v=const$, the incoherent one is assigned to the surfaces $u'=const$.

Now only the factors $u'_n$ and $v^2_q$ have to be separated from each other, and this is clearly done by the fifth condition, according to which the following applies to an oscillator oscillating in a straight line with a single degree of freedom (§\,\ref{Section-17a-5}): 
\vspace*{-3mm}
\begin{align}
u'_n \;=\; n \; \frac{h\:\omega}{\pi} 
     \;=\; g_n \;.\; \frac{\omega}{\pi} 
 \label{label_Planck15c_eq_61}
\;\: ,
\end{align}
i.e. twice the value of $u_n$ in §\,\ref{Section-17a-5}, because the state space contains only positive values of $r$.

Substituting this above gives us: 
\vspace*{0mm}
\begin{align}
4 \: \pi^2 \; v^2_q \;=\; g'_q \;=\; q^2 \; h^2 \; ,
 \quad\quad
v_q \;=\; q \; \frac{h}{2\:\pi} 
 \label{label_Planck15c_eq_62}
\;\: .
\end{align}
This means that for the functions $g$ and $g'$: 
\vspace*{0mm}
\begin{align}
g & \:=\; \frac{\pi \: u'}{\omega} 
  \;=\; \pi \: \left(\: 
         \frac{u}{\omega} \:-\: v 
        \,\right) 
 \label{label_Planck15c_eq_63}
\; , \\
g' & \:=\; 4 \: \pi^2 \; v^2
 \label{label_Planck15c_eq_64}
\; .
\end{align}
If one introduces the semi-axes $a$ and $b<a$ of the orbital ellipse by: 
\vspace*{0mm}
\begin{align}
u \;=\; \frac{m\:\omega^2}{2} \:
        \left(\: a^2 \:+\: b^2 \,\right)
 \quad\quad \mbox{and} \quad\quad
v \;=\; m \; a \; b \; \omega
 \nonumber 
\;\: ,
\end{align}
then the equations 
(\ref{label_Planck15c_eq_61}) and 
(\ref{label_Planck15c_eq_62}) 
for the interfaces of the elementary regions are: 
\vspace*{0mm}
\begin{align}
\left(\: a^2 \:-\: b^2 \,\right)
\;=\; \frac{2\:n\:h}{\pi\:m\:\omega}  
 \quad\quad \mbox{and} \quad\quad
a \; b \;=\; \frac{q\:h}{2\:\pi\:m\:\omega} 
 \label{label_Planck15c_eq_65}
\;\: ,
\end{align}
which can be imagined as symbolized by curves.

These curves become particularly simple if the quantities
\vspace*{0mm}
\begin{align}
\rho^2 \;=\; \left(\frac{a \:+\: b}{2} \,\right)^2
 \quad\quad \mbox{and} \quad\quad
\rho'^2 \;=\; \left(\frac{a \:-\: b}{2} \,\right)^2
 \nonumber 
\;\: 
\end{align}
are chosen for coordinates in the drawing plane, where $\rho$ and $\rho'$ denote the radii of the two circles travelling in opposite directions, into which the oscillation can be broken down, because then you get from (\ref{label_Planck15c_eq_65}) the equations 
\vspace*{0mm}
\begin{align}
\rho'^2 \;=\; \frac{n\:h}{2\:\pi\:m\:\omega}  
   {\color{blue} \:\;=\; 
   \frac{\rho^2}{m\:\omega^2\:\rho^2} 
   \:\left(\frac{n\:h\:\omega}{2\:\pi}\right) 
   }
 \quad\quad \mbox{and} \quad\quad
\rho^2 \;-\; \rho'^2 \;=\; \frac{q\:h}{2\:\pi\:m\:\omega} 
   {\color{blue} \:\;=\; 
   \frac{\rho^2}{2\:m\:\omega^2\:\rho^2} 
   \:\left(\frac{q\:h\:\omega}{\pi}\right) 
   }
 \nonumber 
\;\: 
\end{align}
are obtained for them, which determine the shape of the elementary regions indicated in the 
figure~\ref{fig_Planck_1915c_p449} below.

\begin{figure}[hbt]
\centering
------------------------------------------------------------------------\\
\includegraphics[width=0.5\linewidth]{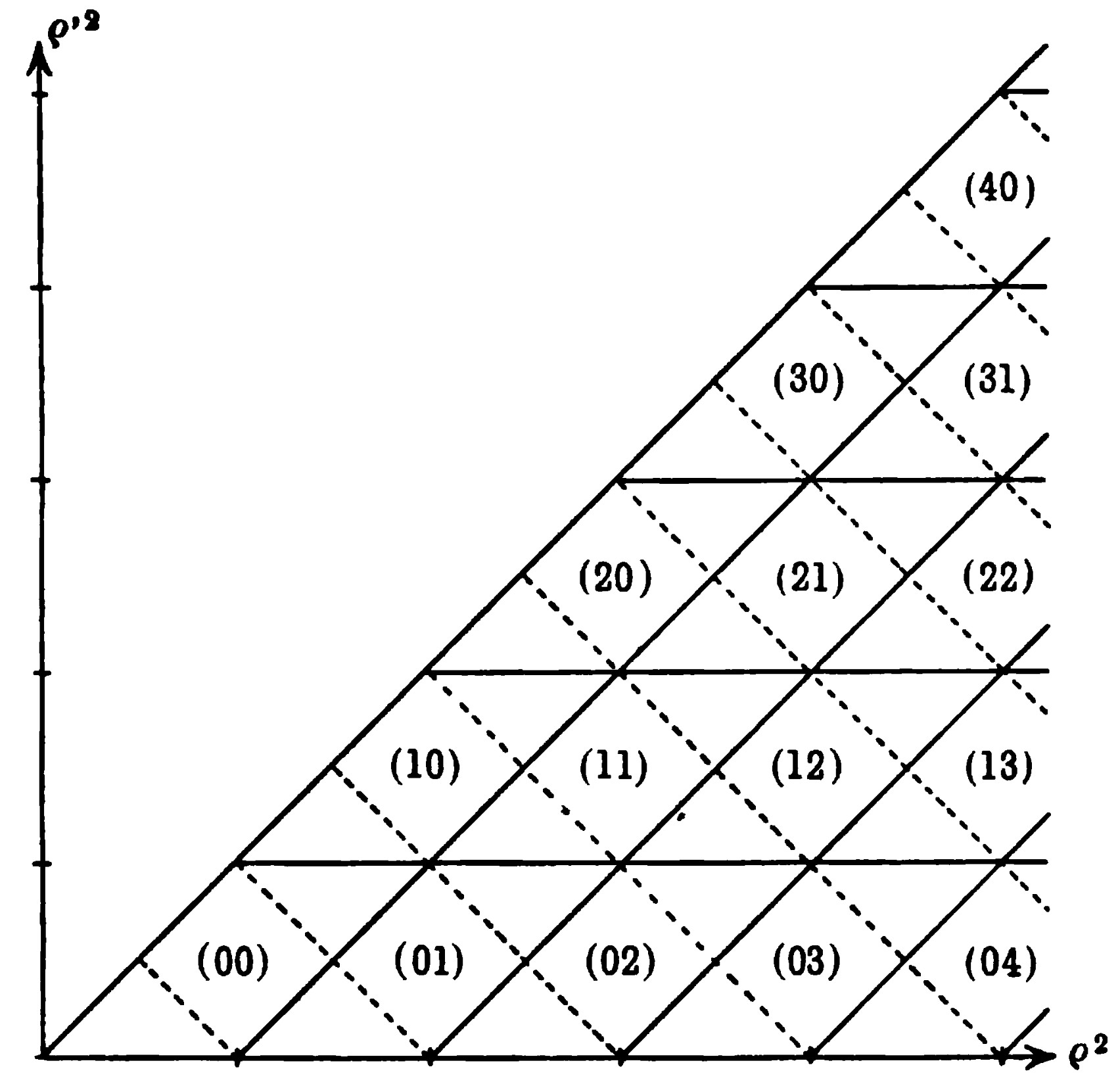} 
\vspace{-3mm}
\caption{\color{black}\it \citet[][p.449.]{Planck_Verh_dtsch_Phys_Ges_1915c}
\label{fig_Planck_1915c_p449}}
------------------------------------------------------------------------
\end{figure}

The distances divided on the coordinate axes correspond to the magnitude $h/(2\:\pi\:m\:\omega)$. 
On one limit line, the abscissa axis $\rho'^2=0$ (axis of circular oscillations), the energy $m\:\omega^2\:\rho^2$ progresses according to whole multiples {\color{blue}(via $n$)}  of $(h\:\omega)/(2\: \pi)$, whereas on the other boundary line, the bisecting line of the coordinate axes $\rho^2=\rho'^2$ (axis of rectilinear oscillations), it progresses according to whole multiples {\color{blue}(via $q$)} of $h\:\omega/\pi$, because there the energy is represented by $2\:m\:\omega^2\:\rho^2$.
The intersection points of the two sets of parallels correspond to certain excellent ellipses whose semi-axes are given by (\ref{label_Planck15c_eq_65}). 
The dashed lines in the figure~\ref{fig_Planck_1915c_p449} below are the lines of constant energy $m\:\omega^2\:\left(\rho^2 \:+\: \rho'^2\right)$, plotted as whole multiples of $(h\:\omega)/(2\: \pi)$.




\section{\underline{** Planck (1915c) -- §\,13  {\it\color{blue}-- Stationary Energy Distributions} } (p.450-451)}
\label{Section-17c-13}
\vspace*{-2mm}

Finally, we calculate the stationary energy distribution over a large number $N$ of the spatially oscillating oscillators under consideration. 
To do this, you first have to calculate the average energy in the elementary region $(n\:q)$ from 
(\ref{label_Planck15c_eq_43}). 
This is done by substituting the values from 
(\ref{label_Planck15c_eq_35}), 
(\ref{label_Planck15c_eq_38}), 
(\ref{label_Planck15c_eq_63}), 
(\ref{label_Planck15c_eq_64})
 with $i=1$ and $i'=2$: 
\vspace*{-3mm}
\begin{align}
g_{\,n} & \:=\; n\; h \; ,
\quad\quad\quad\quad\quad\quad\quad
g'_{\,q}  \;=\; q^2 \; h^2 \; ,
  \nonumber 
  \\
g_{\,n+1} & \:=\; (n\:+\:1)\; h \; ,
\quad\quad\quad
g'_{\,q+1}  \;=\; (q\:+\:1)^2 \; h^2 \; ,
  \nonumber 
  \\
u & \:=\; u' \;+\; \omega\;v 
    \;=\; \frac{\omega}{2\:\pi} \:
       \left(\: 2 \: g \:+\: \sqrt{g'} \:\right)
    \; ,
  \nonumber 
  \\
d\,G & \:=\; dg \;.\; dg' \; ,  
\quad\quad\quad\quad\quad\quad
p \;=\; 2 \: q \;+\; 1 \; ,
  \nonumber 
\end{align}
and results in 
\vspace*{-3mm}
\begin{align}
\overline{u}_{\,n\:q} & \:=\; 
\left(\frac{\omega\:h}{2\:\pi}\right) 
\;.\;
\left(\:
  2\:n \;+\; 1 
  \;+\; \frac{q^2\:+\:q\:+\:1/3}
             {q\:+\:1/2}  
  \:\right) 
\; .
 \label{label_Planck15c_eq_66}
\end{align}
From this follows after (\ref{label_Planck15c_eq_47}) the expression of the characteristic thermodynamic function: 
\vspace*{0mm}
\begin{align}
\Psi & \:=\; N \; k \; 
\ln\!\left[\;
\sum_{n=0}^{\infty} \: 
\sum_{q=0}^{\infty} \: 
\;(\:2\:q\;+\;1\:) \;.\; 
\exp\left( 
-\,\frac{\overline{u}_{\,n\,q}}{k\:T} \right)
\;\right]
\nonumber 
\; , \\
\Psi & \:=\; N \; k \; 
\left\{\:
 -\,\alpha 
 \;-\; \ln\left[\:1\,-\,e^{-\,2\:\alpha}\:\right]
 \;+\;
\ln\!\left[\;
\sum_{q=0}^{\infty} \: 
\;(\,2\,q+1\,) \;.\: 
\exp\left( 
-\,\alpha\:.\:
 \frac{q^2+q+1/3}{q+1/2}
\right)
\;\right]
\:\right\}
\label{label_Planck15c_eq_67}
\; ,
\end{align}
where 
\vspace*{-3mm}
\begin{align}
\alpha \;=\; \frac{h\:\omega}{2\:\pi\;k\;T} \; ,
\nonumber 
\end{align}
\vspace*{0mm}
and according to (\ref{label_Planck15a_eq_21}) the expression of the energy {\it\color{blue}(namely: $U=-\,\partial\,\Psi/\partial \,\tau$ where $\tau=1/T$)}: 
\vspace*{0mm}
\begin{align}
 U & \:=\; \frac{N\:h\:\omega}{2\:\pi} \: 
\left\{\:
 1
 \;+\; 
 \frac{2}{e^{\,2\:\alpha}-1}
 \;+\;
\frac{\displaystyle
\sum_{q=0}^{\infty} \: 
\;(\,q^2+q+1/3\,) \;.\: 
\exp\left( 
-\,\alpha\:.\:
 \frac{q^2+q+1/3}{q+1/2}
\right)
}{\displaystyle
\sum_{q=0}^{\infty} \: 
\;(\,q+1/2\,) \;.\: 
\exp\left( 
-\,\alpha\:.\:
 \frac{q^2+q+1/3}{q+1/2}
\right)
}
\:\right\}
\label{label_Planck15c_eq_68}
\; .
\end{align}

For \dashuline{high temperatures} $(\alpha \ll 1, \:q \gg 1)$ one can replace the summations by integrations and obtain: 
\vspace*{0mm}
\begin{align}
\Psi & \:=\; N \; k \; 
\left\{\:
 \;-\; \ln\left(\:2\:\alpha\:\right)
 \;+\;
\ln\!\left[\;
\bigintsss_{\;0}^{\,\infty}
\,2\,q \;.\: e^{\,-\,\alpha\:q} \;.\: dq
\;\right]
\:\right\}
\nonumber 
\; , \\
\Psi & \:=\; -\:3 \; N \; k \;  
             \ln\left(\,\alpha\,\right)
{\color{blue}\:
\;=\; -\:3 \; N \; k \;  
             \ln\left(\,
             \frac{h\:\omega}{2\:\pi\;k\;T} 
             \,\right)
}
\nonumber 
\; .
\end{align}
From this, according to (\ref{label_Planck15a_eq_21}) {\it\color{blue}and $\:U=T^2\,\partial\,\Psi/\partial \,T$,} 
the energy 
\vspace*{0mm}
\begin{align}
U & \:=\; 3 \; N \; k \;  T
\nonumber 
\; , 
\end{align}
corresponds to the three degrees of freedom.

For \dashuline{extremely low temperatures} $(\alpha \gg 1, \: n=0, \: q=0)$ the {\bf\dashuline{zero point energy}} results from 
(\ref{label_Planck15c_eq_68}) or 
(\ref{label_Planck15c_eq_66}) as: 
\vspace*{-3mm}
\begin{align}
U & \:=\; \frac{N\:h\:\omega}{2\:\pi} \;.\; \frac{5}{3} 
{\color{blue} \: \:\;=\;
    N\;.\,
    \left[\:\frac{h\:\omega}{4\:\pi}\:\right] 
    \;.\, 
    \left( \, 3 \:+\: \frac{1}{3} \, \right)
}
\label{label_Planck15c_eq_69}
\; , 
\end{align}
i.e. not three times, but three and a third times the {\bf\dashuline{zero point energy}} of rectilinear 
oscillators{\color{blue}$\,$\footnote{$\:${\it\color{blue}Namely $10/3$ times $E_0$ computed with $n=0$ in the relationship (\ref{label_Planck15a_eq_24bis}) or $E_n = (n+1/2)\:.\: \hslash\:\omega = (2\:n+1)\:.\: (h\:\omega)/(4\:\pi)$ / P. Marquet}}}.

\vspace*{-4mm}
\begin{center}
---------------------------------------------------
\end{center}
\vspace*{-3mm}

If one assumes that the oscillations of the oscillators are not spatial {\color{blue}(i.e. $f \neq 3$)}, but rather presuppose $f=2$, then the theory developed here for the structure of the state space results in exactly the same laws as those symbolized in the figure above, but through a much simpler calculation.
For the energy of a system of $N$ of such plane oscillators, the expression:
\vspace*{0mm}
\begin{align}
{\color{blue} U \;=\;\:} 
\frac{N\:h\:\omega}{2\:\pi} \:
\left(\;
\frac{3}{2} 
\:+\: 
\frac{1}{\exp(\alpha) \:-\: 1 } 
\:+\: 
\frac{2}{\exp(2\:\alpha) \:-\: 1 } 
\;\right)
\; 
\label{label_Planck15c_eq_70}
\end{align}
takes the place of (\ref{label_Planck15c_eq_68}).

Here the {\bf\dashuline{zero point energy}} is just three times that of a rectilinear oscillator, while for high temperatures the energy takes on the value $\,2\:N\:k\:T\,$ determined by the two degrees of freedom.


\vspace*{-4mm}
\begin{center}
---------------------------------------------------
\end{center}
\vspace*{-3mm}

\newpage
\setcounter{section}{15} 
\section{\underline{\color{blue}*** The German Papers of Planck (1915a,b)}}
\label{Section-PDF-1915a-1915b}
\vspace*{-6mm}
\begin{figure}[hbt]
\centering
\includegraphics[width=0.65\linewidth]{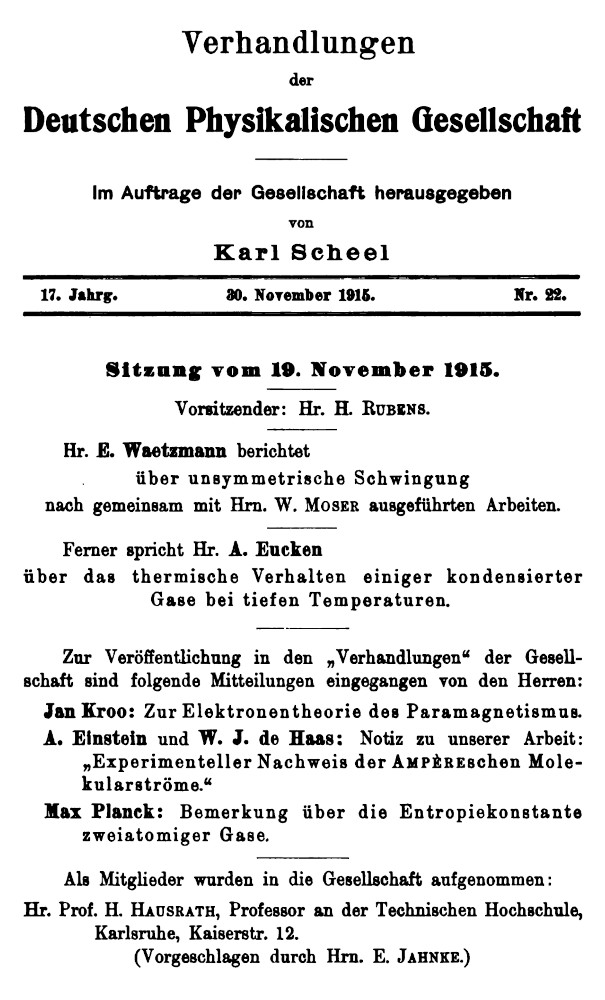}
\end{figure}
\clearpage
\begin{figure}[hbt]
\centering
\includegraphics[width=0.8\linewidth]{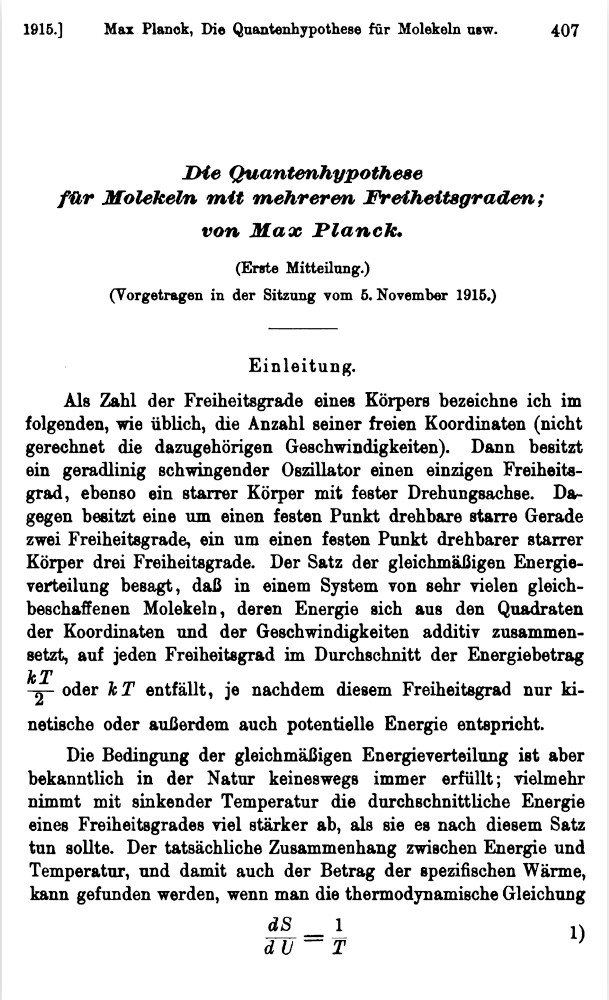}
\end{figure}
\begin{figure}[hbt]
\centering
\includegraphics[width=0.8\linewidth]{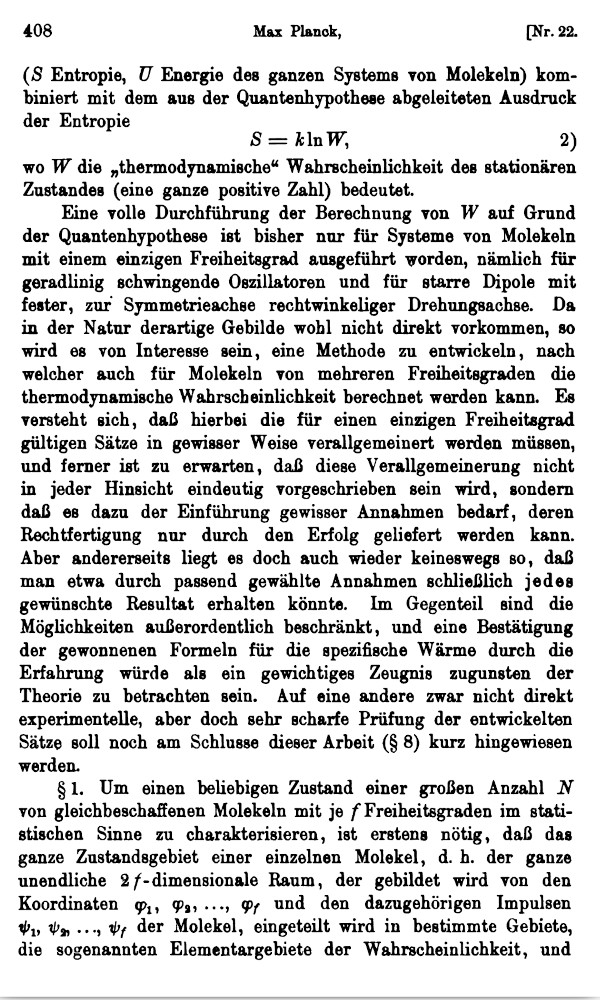}
\end{figure}
\begin{figure}[hbt]
\centering
\includegraphics[width=0.8\linewidth]{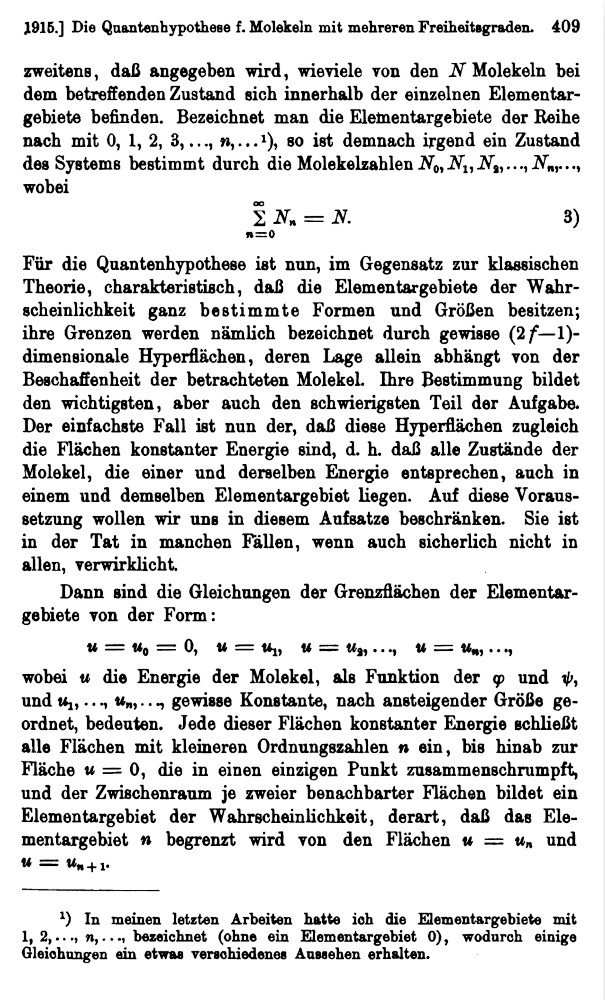}
\end{figure}
\begin{figure}[hbt]
\centering
\includegraphics[width=0.8\linewidth]{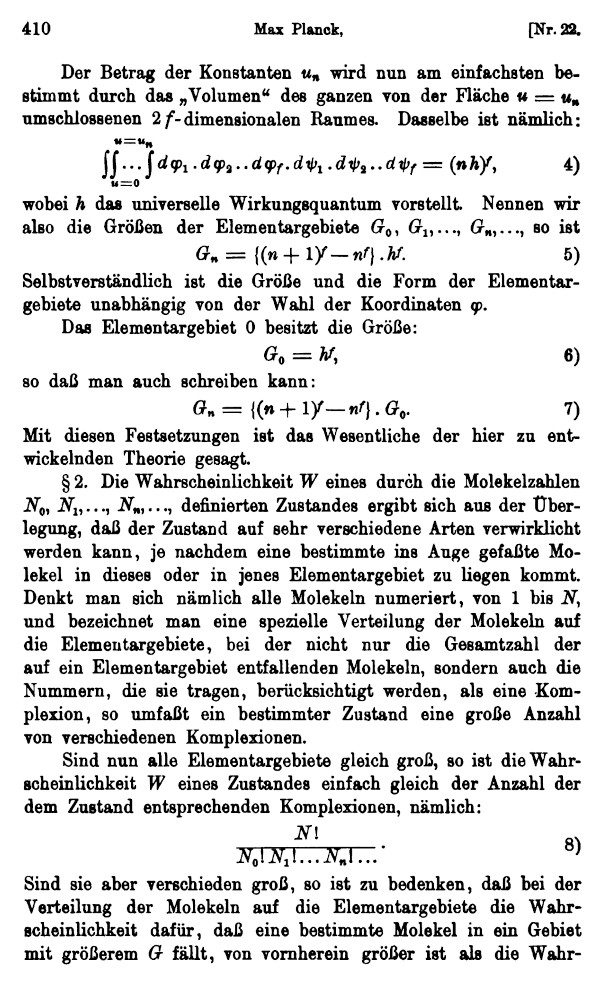}
\end{figure}
\begin{figure}[hbt]
\centering
\includegraphics[width=0.8\linewidth]{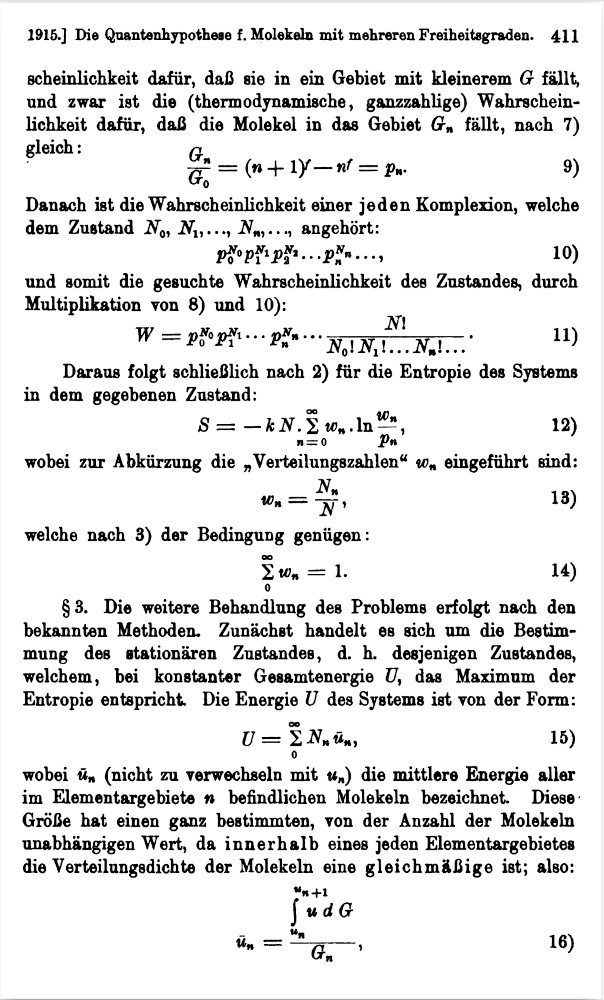}
\end{figure}
\begin{figure}[hbt]
\centering
\includegraphics[width=0.8\linewidth]{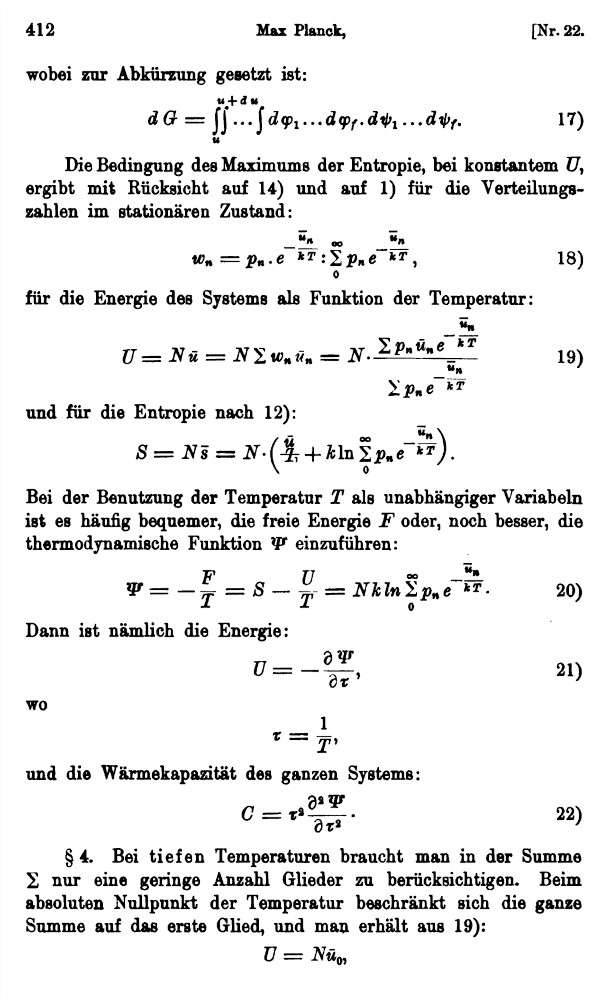}
\end{figure}
\begin{figure}[hbt]
\centering
\includegraphics[width=0.8\linewidth]{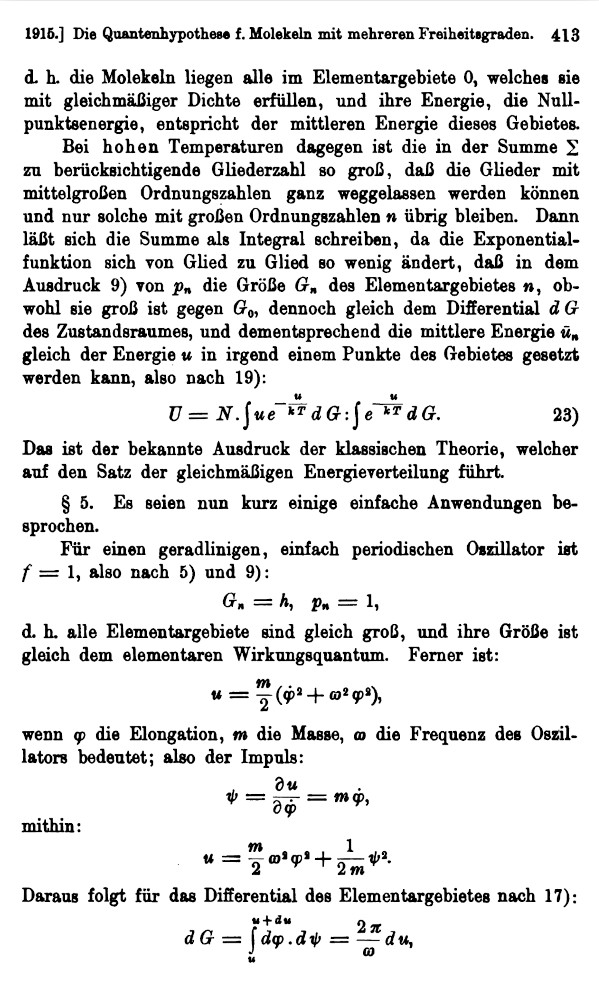}
\end{figure}
\begin{figure}[hbt]
\centering
\includegraphics[width=0.8\linewidth]{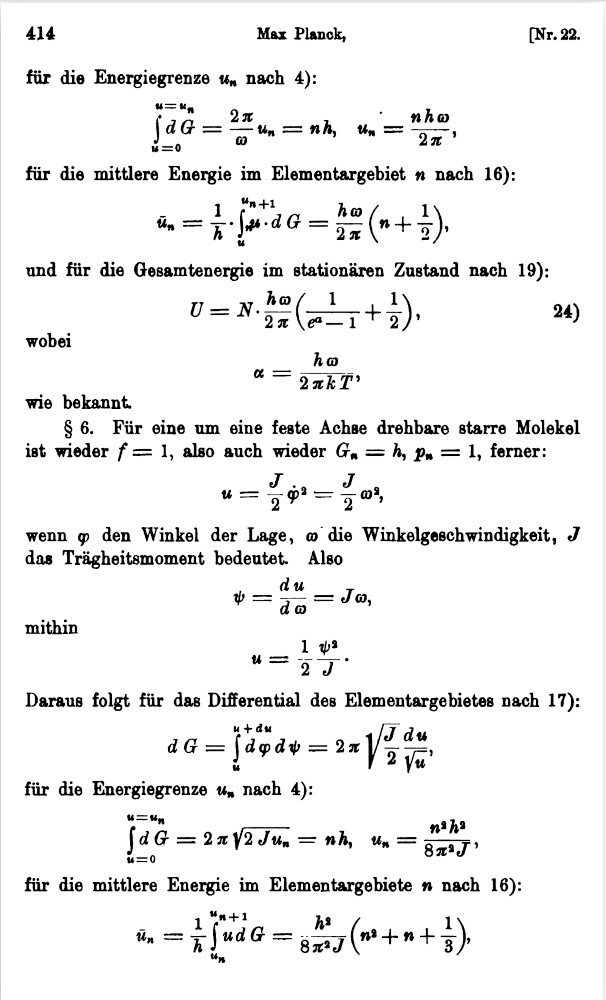}
\end{figure}
\begin{figure}[hbt]
\centering
\includegraphics[width=0.8\linewidth]{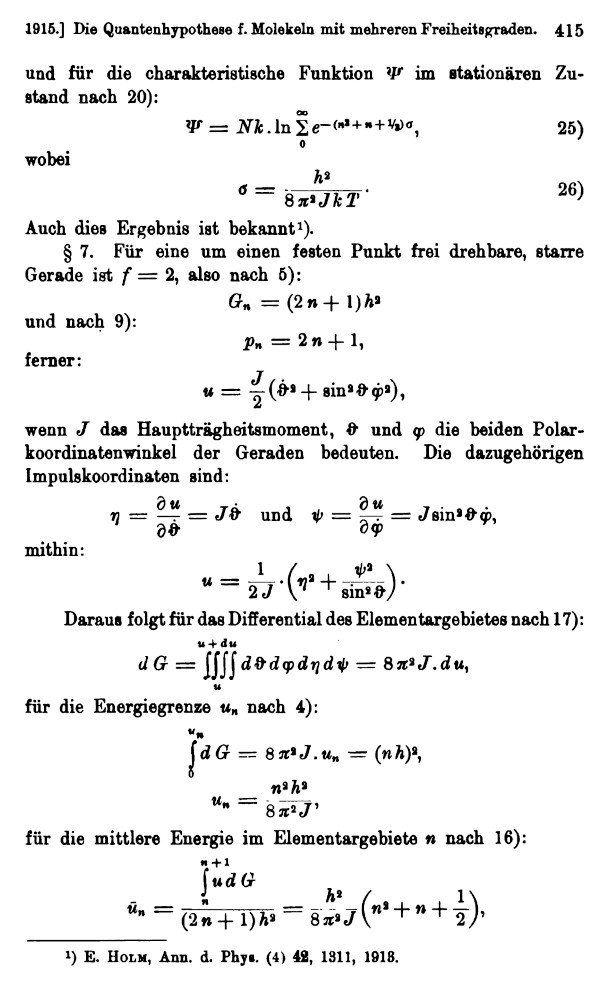}
\end{figure}
\begin{figure}[hbt]
\centering
\includegraphics[width=0.8\linewidth]{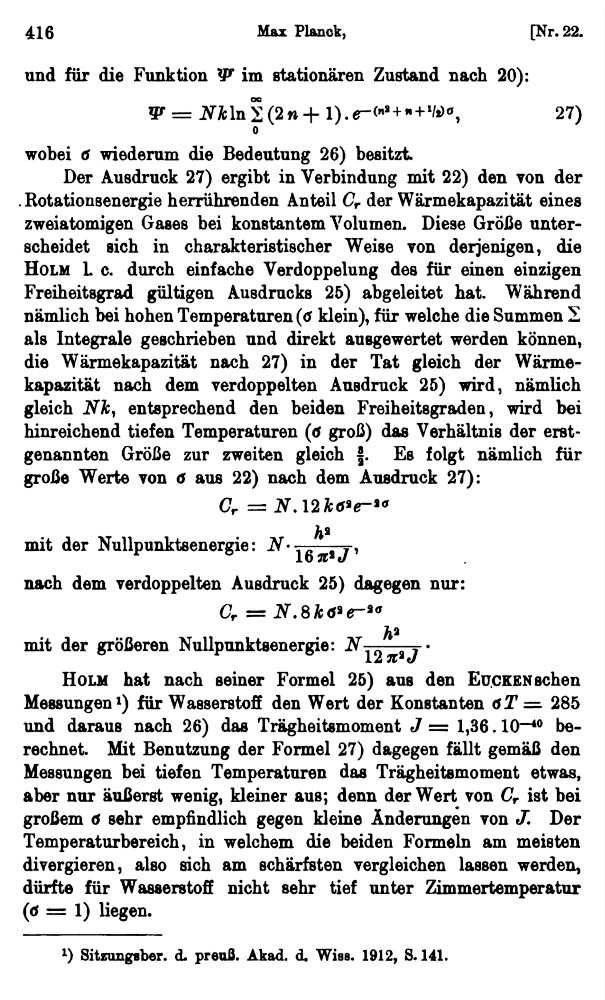}
\end{figure}
\begin{figure}[hbt]
\centering
\includegraphics[width=0.8\linewidth]{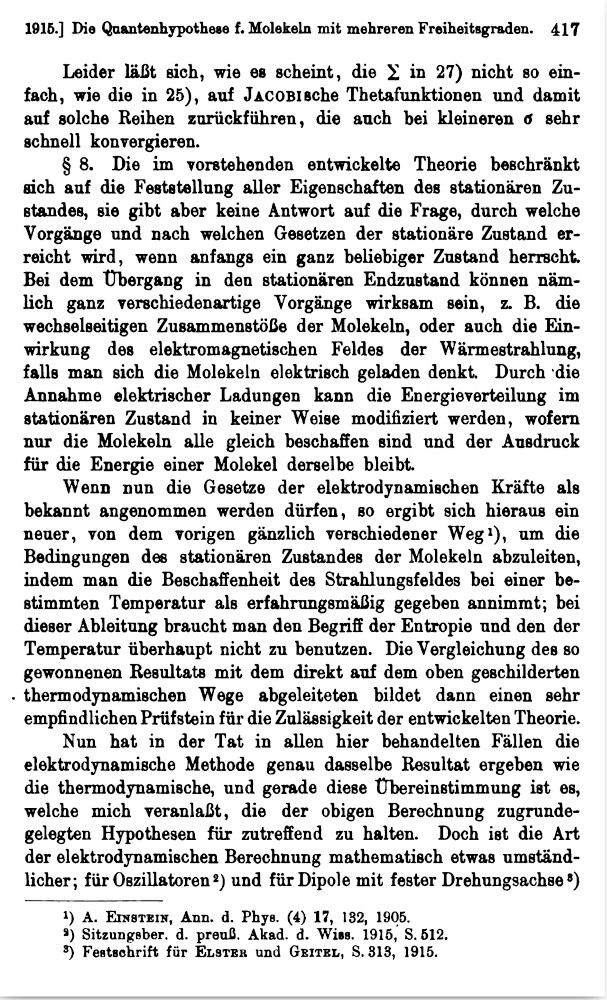}
\end{figure}
\begin{figure}[hbt]
\centering
\includegraphics[width=0.8\linewidth]{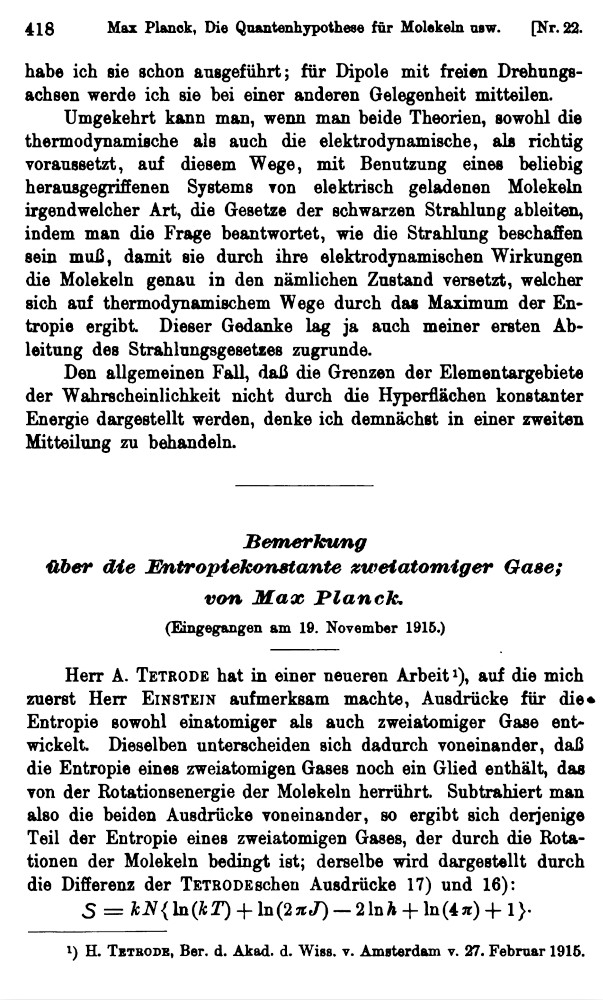}
\end{figure}
\begin{figure}[hbt]
\centering
\includegraphics[width=0.8\linewidth]{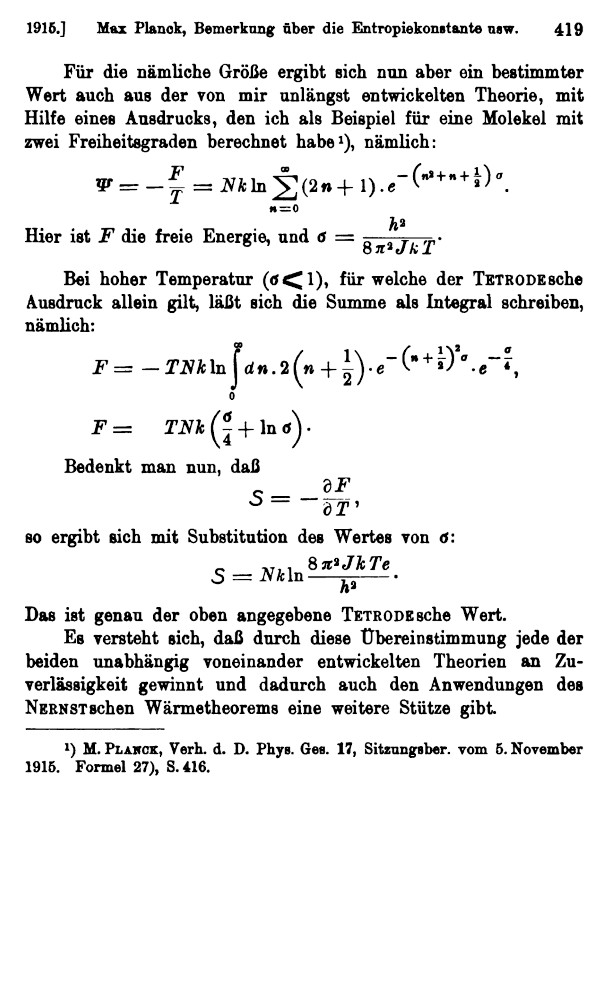}
\end{figure}
\clearpage

\newpage
\setcounter{section}{16} 
\section{\underline{\color{blue}**** The German Paper of Planck (1915c)}}
\label{Section-PDF-1915c}
\vspace*{-4mm}
\begin{figure}[hbt]
\centering
\includegraphics[width=0.65\linewidth]{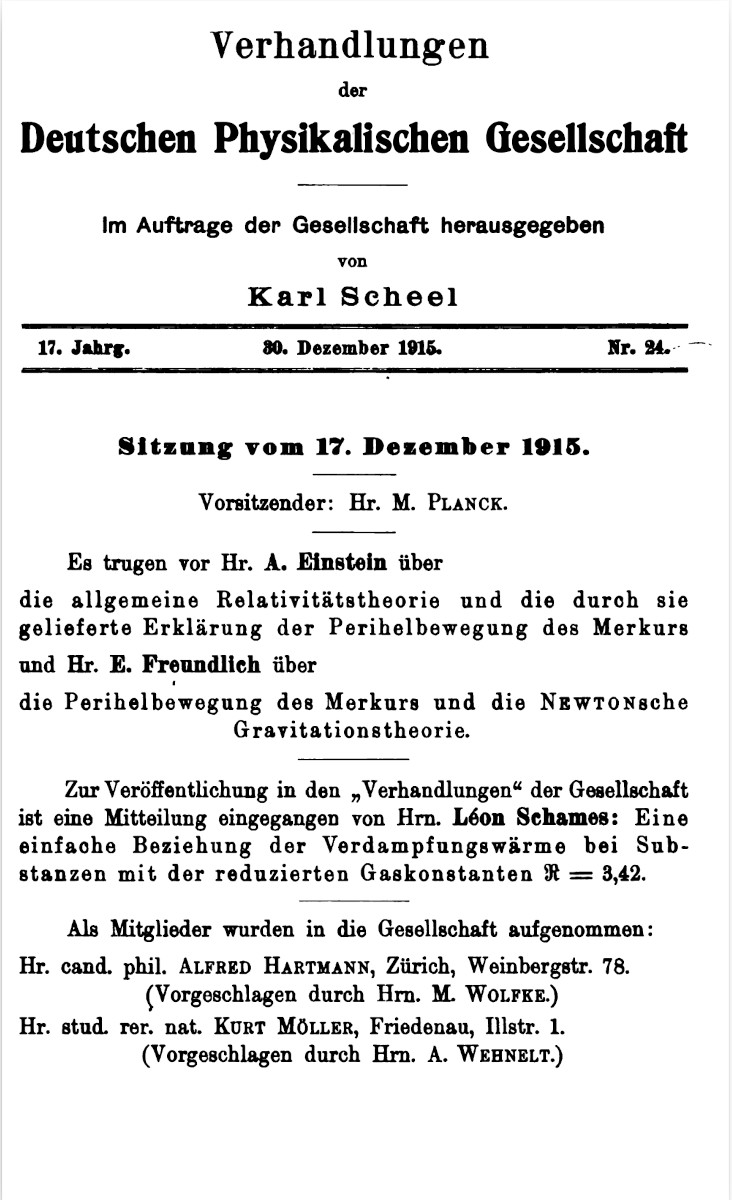}
\end{figure}
\clearpage
\begin{figure}[hbt]
\centering
\includegraphics[width=0.8\linewidth]{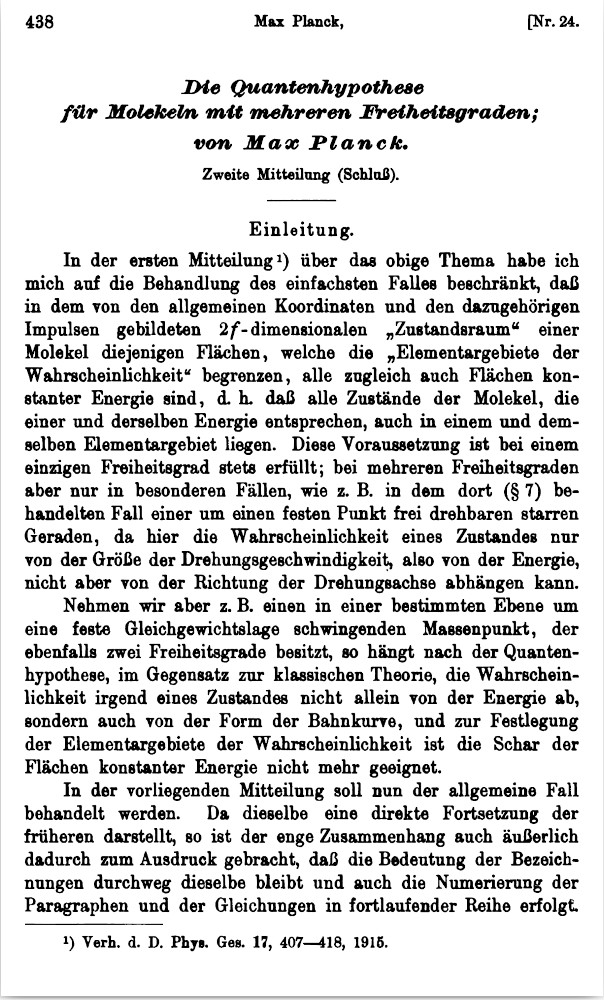}
\end{figure}
\begin{figure}[hbt]
\centering
\includegraphics[width=0.8\linewidth]{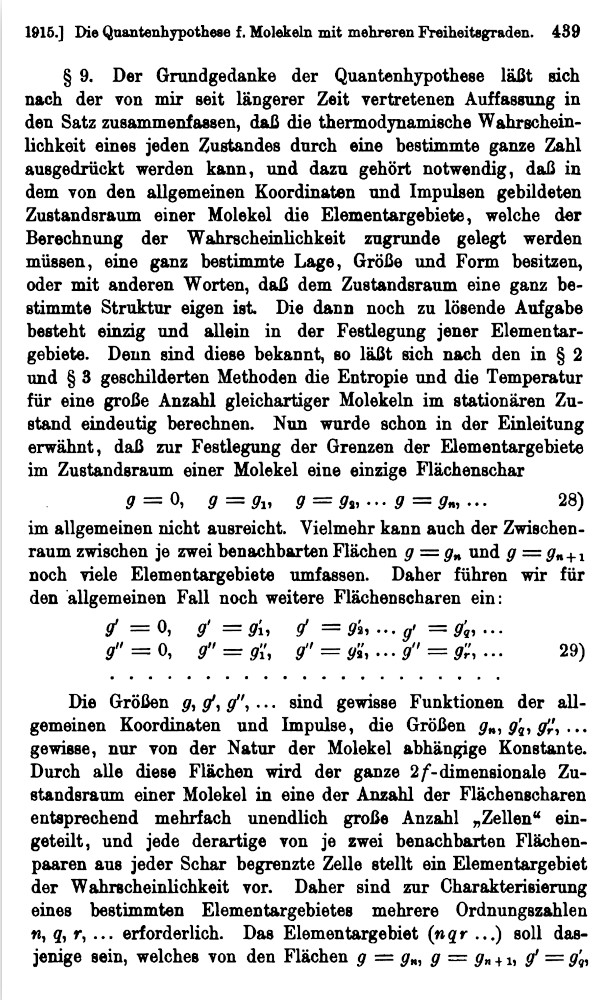}
\end{figure}
\begin{figure}[hbt]
\centering
\includegraphics[width=0.8\linewidth]{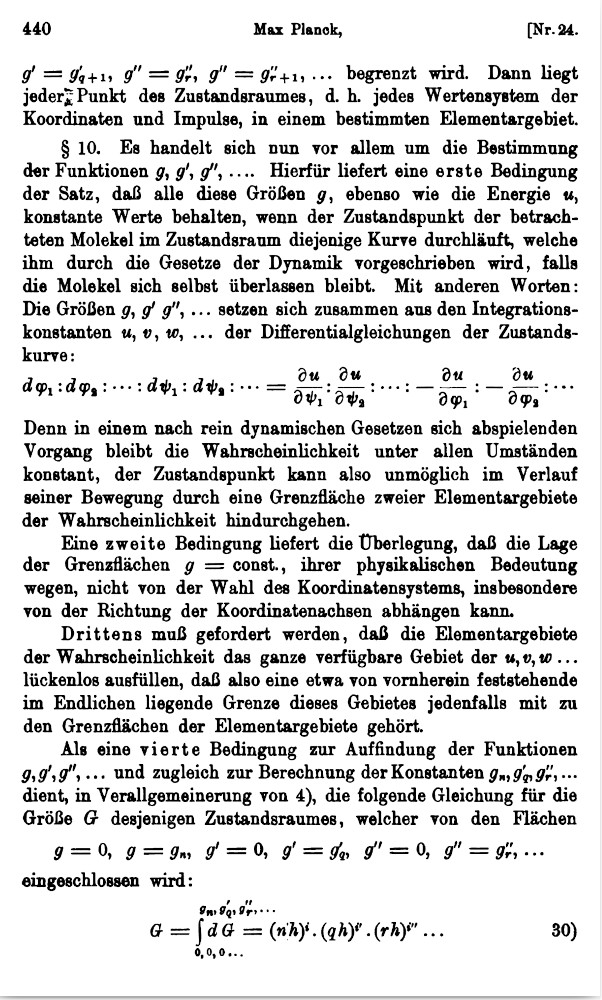}
\end{figure}
\begin{figure}[hbt]
\centering
\includegraphics[width=0.8\linewidth]{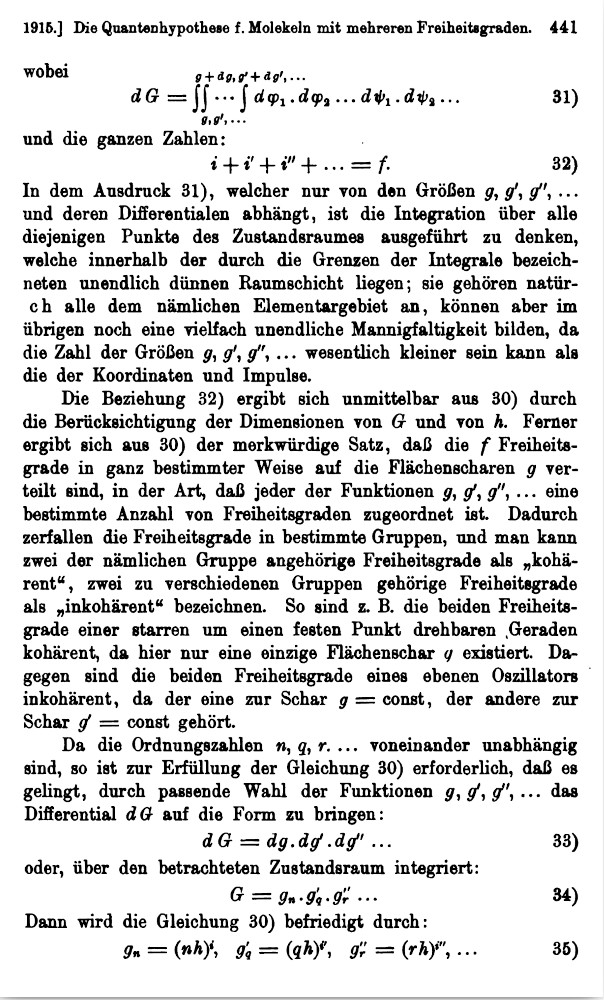}
\end{figure}
\begin{figure}[hbt]
\centering
\includegraphics[width=0.8\linewidth]{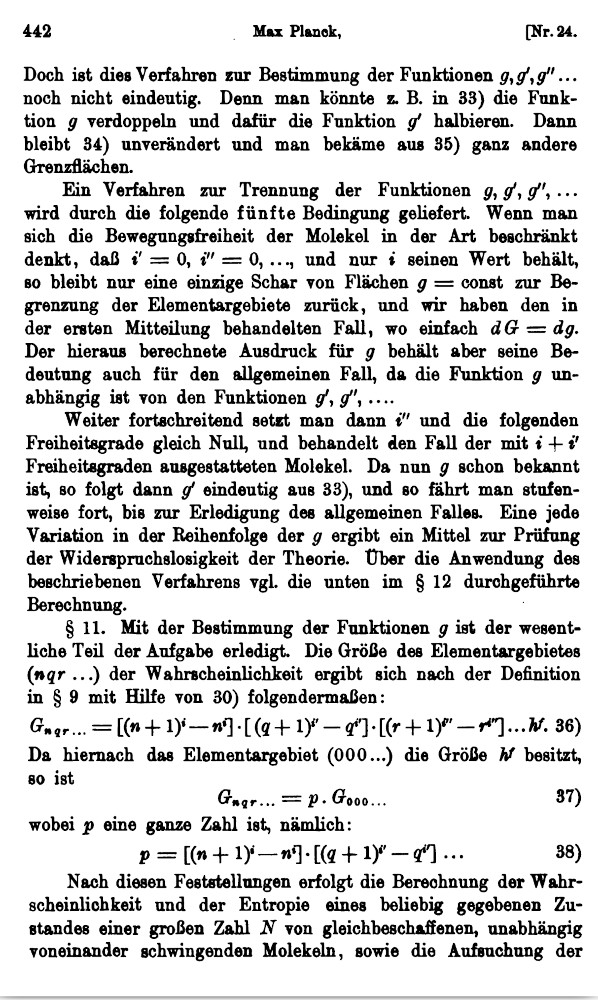}
\end{figure}
\begin{figure}[hbt]
\centering
\includegraphics[width=0.8\linewidth]{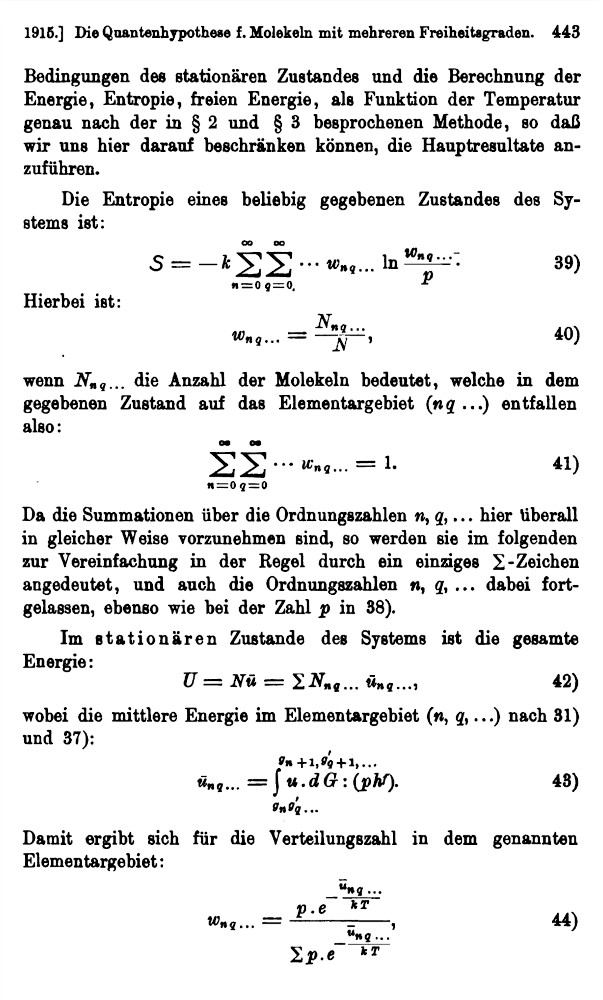}
\end{figure}
\begin{figure}[hbt]
\centering
\includegraphics[width=0.8\linewidth]{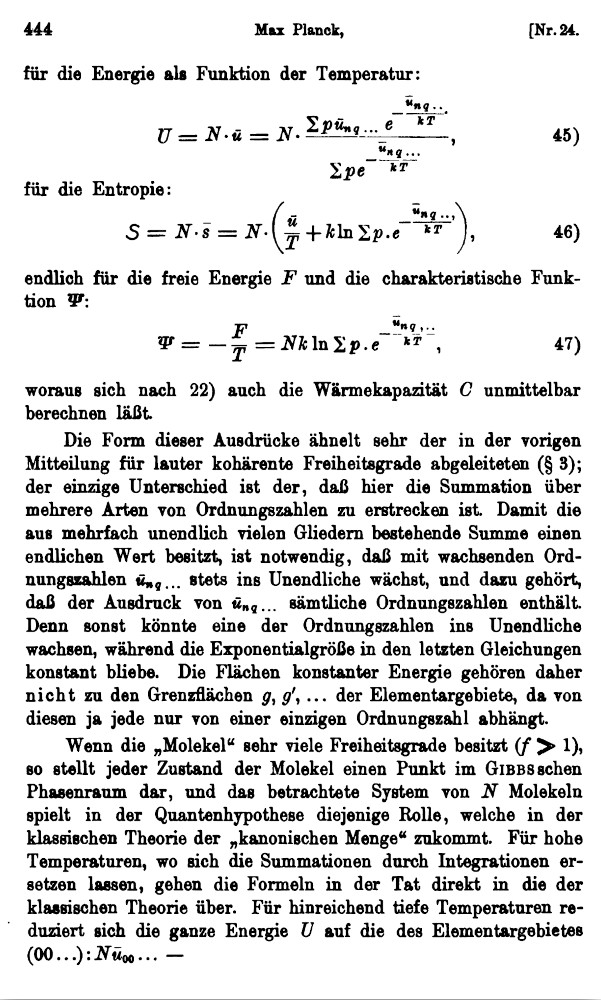}
\end{figure}
\begin{figure}[hbt]
\centering
\includegraphics[width=0.8\linewidth]{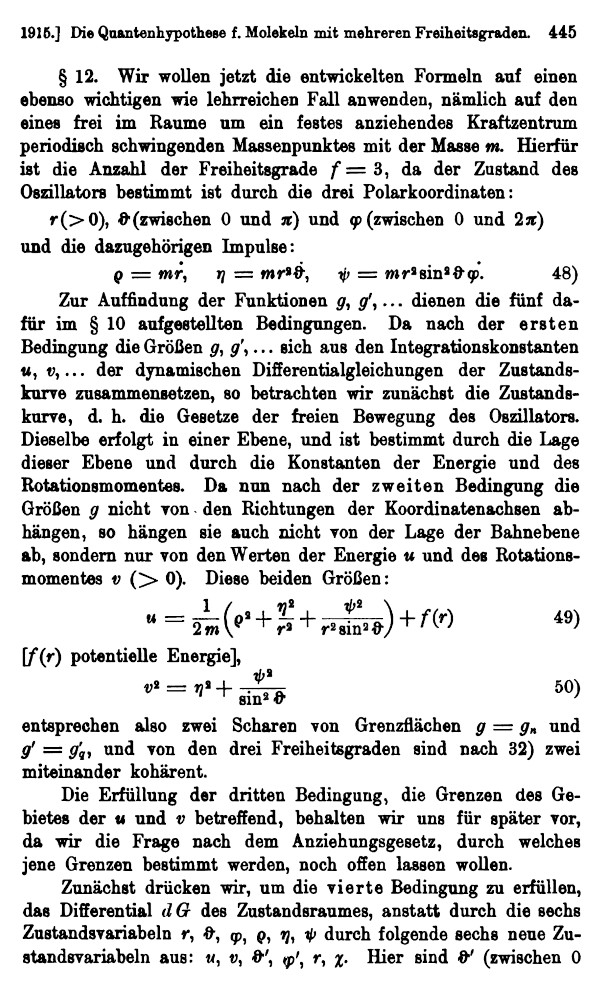}
\end{figure}
\begin{figure}[hbt]
\centering
\includegraphics[width=0.8\linewidth]{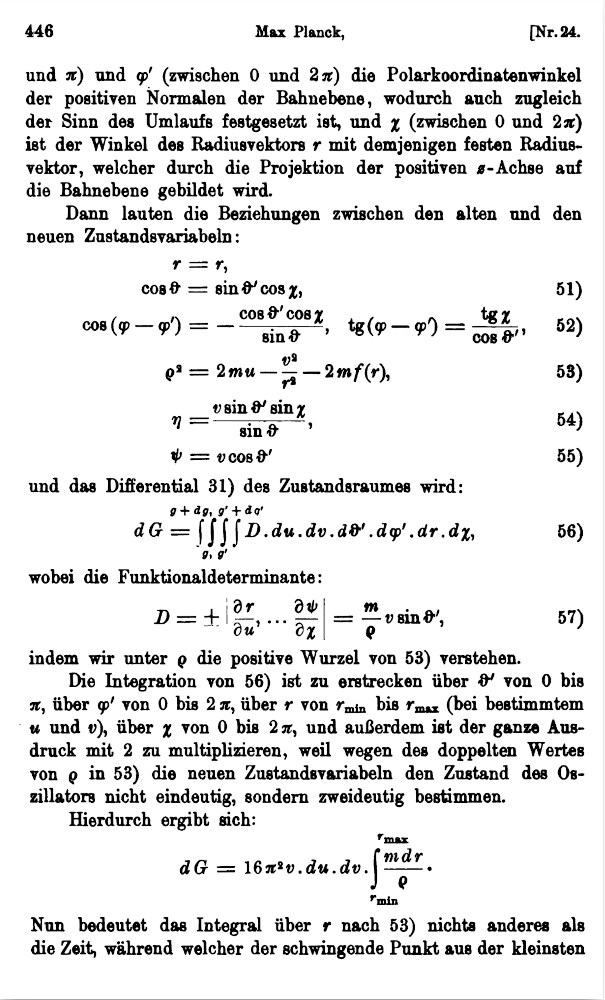}
\end{figure}
\begin{figure}[hbt]
\centering
\includegraphics[width=0.8\linewidth]{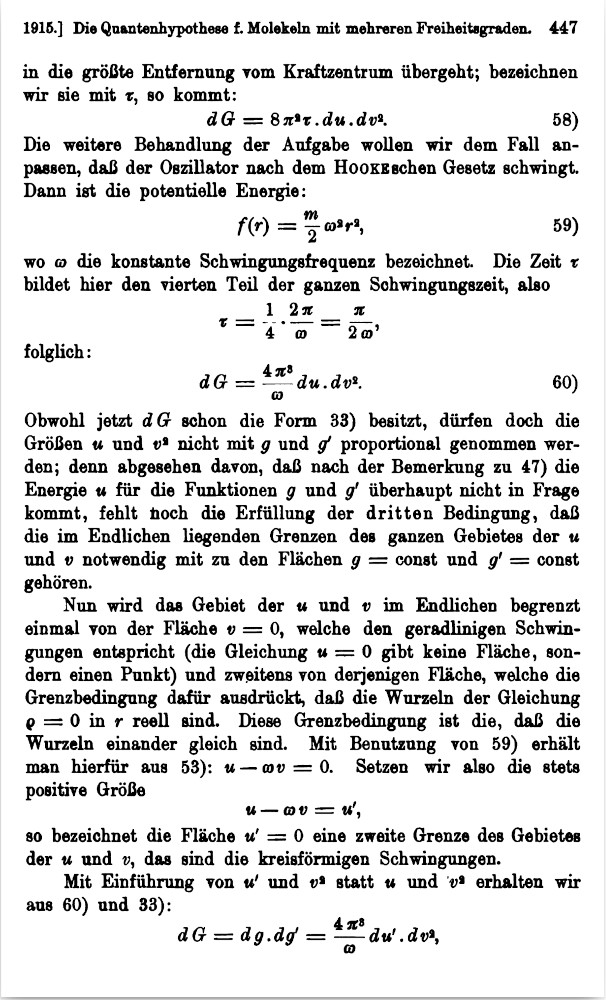}
\end{figure}
\begin{figure}[hbt]
\centering
\includegraphics[width=0.8\linewidth]{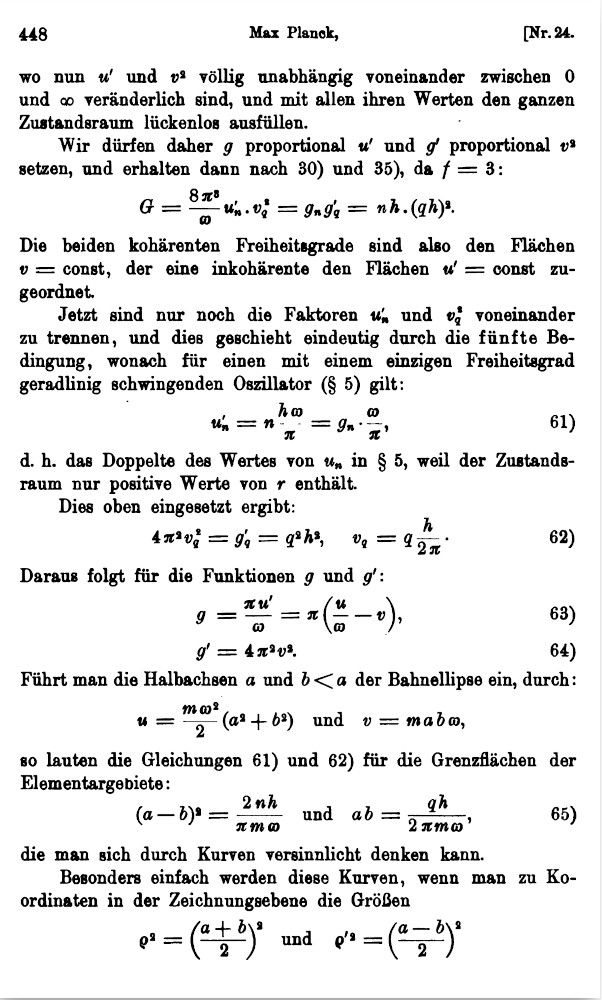}
\end{figure}
\begin{figure}[hbt]
\centering
\includegraphics[width=0.8\linewidth]{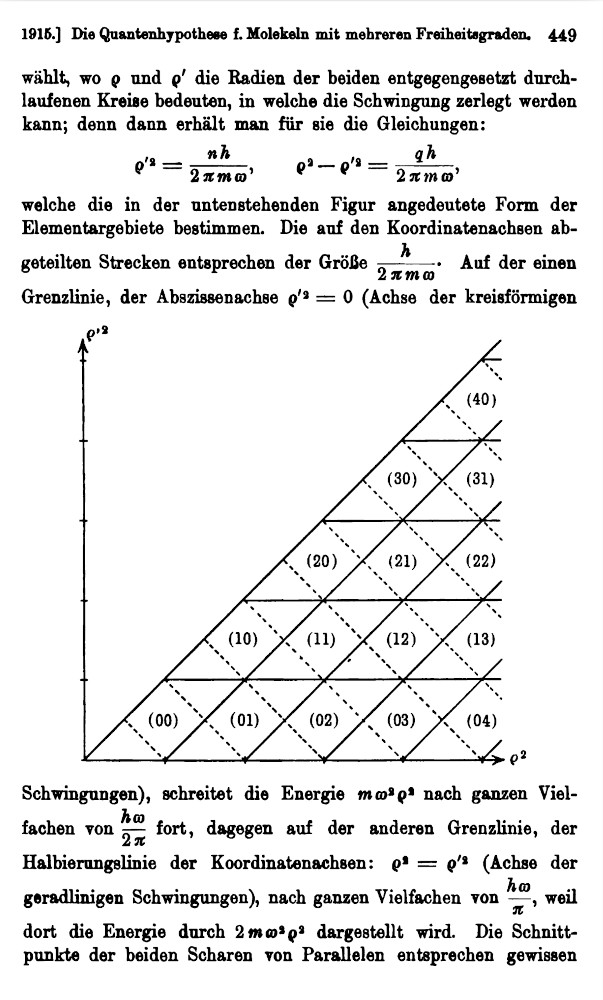}
\end{figure}
\begin{figure}[hbt]
\centering
\includegraphics[width=0.8\linewidth]{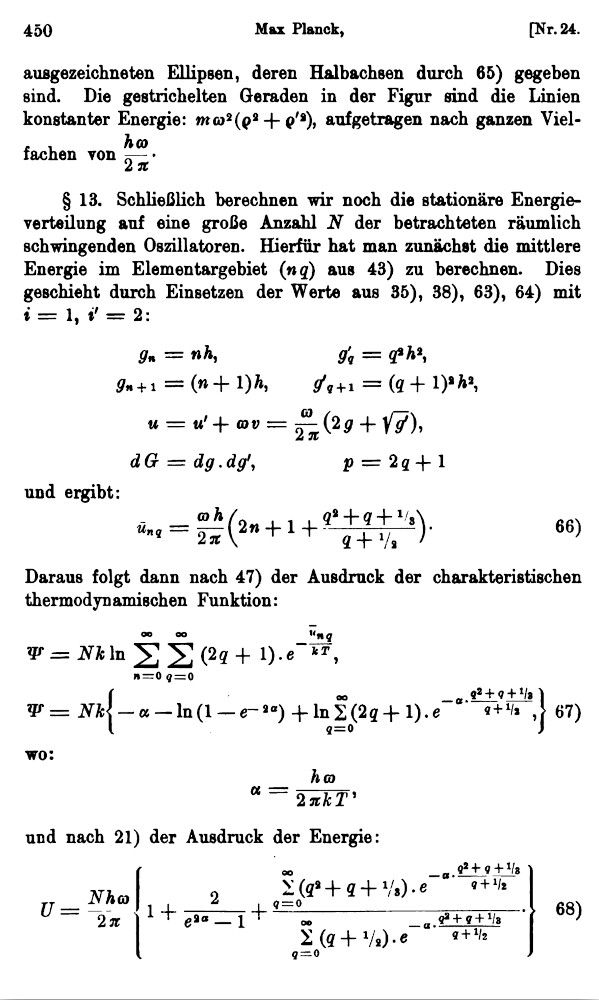}
\end{figure}
\begin{figure}[hbt]
\centering
\includegraphics[width=0.8\linewidth]{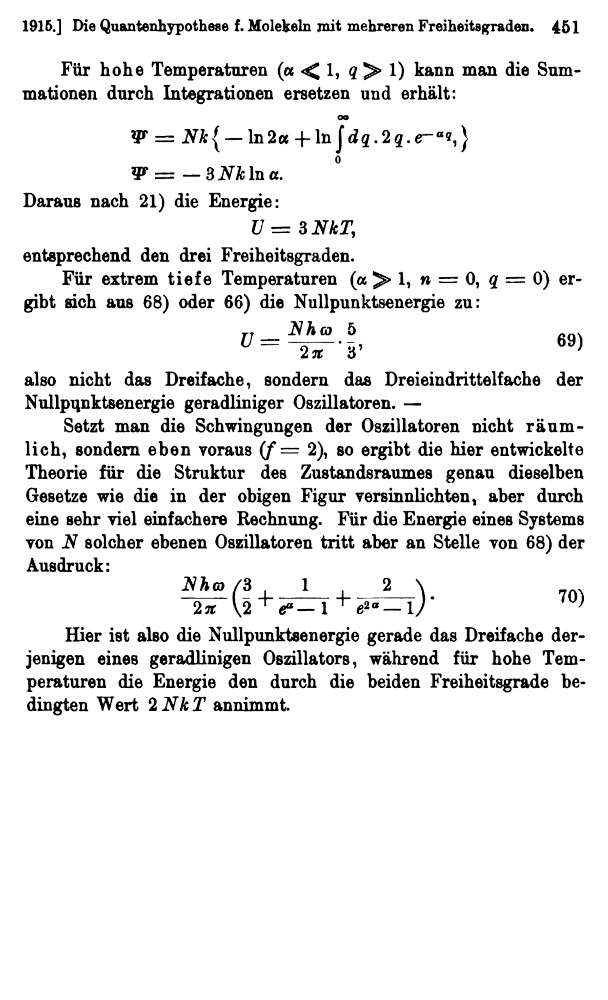}
\end{figure}

\end{document}